\documentclass[11pt]{article} 
\pdfoutput=1
\usepackage{amssymb}
\usepackage{amsmath}
\usepackage{natbib}
\usepackage{graphicx}
\usepackage{algorithmic}
\usepackage{algorithm}
\usepackage{subfigure}
\usepackage{amsthm}
\usepackage{epstopdf}

\usepackage[margin=3cm]{geometry}

\begin{document}

\newcommand{\vect}[1]{\boldsymbol{#1}}

\setlength{\parindent}{0pc}
\setlength{\parskip}{1ex}


\title{Approximating the Likelihood in Approximate Bayesian Computation} 
\author{Christopher C Drovandi*, Clara Grazian$\dagger$, Kerrie Mengersen*, Christian Robert$\ddagger$ \\ \\ *School of Mathematical Sciences, Queensland University of Technology, Australia \\ \\
$\dagger$Nuffield Department of Medicine, University of Oxford, England \\ \\
$\ddagger$ Universit\'e Paris-Dauphine, PSL, France \\  Department of Statistics, University of Warwick, England \\ \\
email: \texttt{c.drovandi@qut.edu.au}
}
\maketitle

ABSTRACT

The conceptual and methodological framework that underpins approximate Bayesian computation (ABC) is targetted primarily towards problems in which the likelihood is either challenging or missing.  ABC uses a simulation-based non-parametric estimate of the likelihood of a summary statistic and assumes that the generation of data from the model is computationally cheap.  This chapter reviews two alternative approaches for estimating the intractable likelihood, with the goal of reducing the necessary model simulations to produce an approximate posterior.  The first of these is a Bayesian version of the synthetic likelihood (SL), initially developed by \citet{Wood2010}, which uses a multivariate normal approximation to the summary statistic likelihood.  Using the parametric approximation as opposed to the non-parametric approximation of ABC, it is possible to reduce the number of model simulations required.  The second likelihood approximation method we consider in this chapter is based on the empirical likelihood (EL), which is a non-parametric technique and involves maximising a likelihood constructed empirically under a set of moment constraints.  \citet{Mengersen2013} adapt the EL framework so that it can be used to form an approximate posterior for problems where ABC can be applied, that is, for models with intractable likelihoods.  However, unlike ABC and the Bayesian SL (BSL), the Bayesian EL ($BC_{el}$) approach can be used to completely avoid model simulations in some cases.  The BSL and BCel methods are illustrated on models of varying complexity.   


\vspace{0.5cm}

KEYWORDS:\\
Approximate Bayesian computation, empirical likelihood, importance sampling, $BC_{el}$, sequential Monte Carlo, synthetic likelihood 

\vspace{0.5cm}


\section{Introduction}

Approximate Bayesian computation (ABC) is now a mature algorithm for likelihood-free estimation. It has been successfully applied to a wide range of real-world problems for which more standard analytic tools were unsuitable due to the absence or complexity of the associated likelihood. It has also paved the way for a range of algorithmic extensions that take advantage of appealing ideas embedded in other approaches.   Despite the usefulness of ABC, the method does have a number of drawbacks.  The approach is simulation intensive, requires tuning of the tolerance threshold, discrepancy function and weighting function, and suffers from a curse of dimensionality of the summary statistic.  The latter issue stems from the fact that ABC uses a non-parametric estimate of the likelihood function of a summary statistic \citep{Blum2009}.

In this chapter we review two alternative methods of approximating the intractable likelihood function for the model of interest, both of which aim to improve computational efficiency relative to ABC.  The first is the synthetic likelihood (SL, originally developed by \citet{Wood2010}), which uses a multivariate normal approximation to the summary statistic likelihood.  This auxiliary likelihood can be maximised directly or incorporated in a Bayesian framework, which we refer to as BSL.    The BSL approach requires substantially less tuning than ABC.  Further, BSL scales more efficiently with an increase in the dimension of the summary statistic due to the parametric approximation of the summary statistic likelihood.  However, the BSL approach remains simulation intensive.  In another chapter, \citet{Fasiolo2014} apply BSL to dynamic ecological models and compare it with an alternative Bayesian method for state space models.  In this chapter, we provide a more thorough review of SL both in the classical and Bayesian frameworks.  

The second approach we consider uses an empirical likelihood (EL) within a Bayesian framework ($BC_{el}$, see \citet{Mengersen2013}).   This approach can in some cases avoid the need for model simulation completely and inherits the established theoretical and practical advantages of synthetic likelihood.  This improvement in computational efficiency is at the expense  of specification of constraints and making equivalence statements about parameters under the different models. Of note is that the latter enables, for the first time, model comparison using Bayes factors even if the priors are improper. In summary, in the Bayesian context, both of these approaches replace intractable likelihoods with alternative likelihoods that are more manageable computationally.



\section{Synthetic Likelihood}

The first approach to approximating the likelihood that is considered in this chapter is the use of a synthetic likelihood (SL), which was first introduced by \citet{Wood2010}.  The key idea behind the SL is the assumption that the summary statistic conditional on a parameter value has a multivariate normal distribution with mean vector $\mu(\theta)$ and covariance matrix $\Sigma(\theta)$.  That is, we assume that
\begin{align*}
p(s|\theta) &= {\cal N}(s;\mu(\theta),\Sigma(\theta)),
\end{align*}
where ${\cal N}$ denotes the density of the multivariate normal distribution.  Of course, in general for models with intractable likelihoods, the distribution of the summary statistic is unknown and thus $\mu(\theta)$ and $\Sigma(\theta)$ are generally unavailable.  However, it is possible to estimate these quantities empirically via simulation.  Consider generating $n$ independent and identically distributed (iid) summary statistic values, $s^{1:n} = (s^1,\ldots,s^n)$, from the model based on a particular value of $\theta$, $s^{1:n} \stackrel{\mathrm{iid}}{\sim} p(s|\theta)$.  Then the mean and covariance matrix can be estimated via
\begin{align}
\begin{split}
\mu(\theta) &\approx \mu_n(\theta) = \frac{1}{n}\sum_{i=1}^n s^i, \\
\Sigma(\theta) &\approx \Sigma_n(\theta) = \frac{1}{n-1}\sum_{i=1}^n (s^i - \mu_n(\theta))(s^i - \mu_n(\theta))^\top, \label{eq:mean_cov_estimates}
\end{split}
\end{align}
where the superscript $\top$ denotes transpose.  The likelihood of the observed summary statistic, $s_{obs}$, is estimated via $p_n(s_{obs}|\theta) = {\cal N}(s_{obs};\mu_n(\theta),\Sigma_n(\theta))$.  We use the subscript $n$ on $p_n(s_{obs}|\theta)$ to denote the fact that the approximate likelihood will depend on the choice of $n$.  The larger the value of $n$ the better the mean and covariance parameters of the multivariate normal distribution can be approximated.  However, larger values of $n$ need more computation to estimate the likelihood.  It is likely that a suitable value of $n$ will be problem dependent, in particular, it may depend on the actual distribution of the summary statistic and also the dimension of the summary statistic.  The value of $n$ must be large enough so that the empirical covariance matrix is positive definite.  

Note that \citet{Wood2010} described some extensions, such as using robust covariance matrix estimation to handle some non-normality in the summary statistics and robustifying the SL when the observed summary statistic falls in the tails of the summary statistic distribution (i.e.\ when a poor parameter value is proposed or when the model is mis-specified).

The SL may be incorporated into a classical or Bayesian framework, which are both described below.  Then, attempts in the literature to accelerate the SL method are described.  We finish the section with a real data example in cell biology.

\subsection{Classical synthetic Likelihood}

The approach adopted in \citet{Wood2010} is to consider the following estimator
\begin{align}
\hat{\theta}_n &= \arg \max_{\theta} {\cal N}(s_{obs};\mu_n(\theta),\Sigma_n(\theta)), \label{eq:max_sl}
\end{align}
which is the maximum SL estimator.  We use the subscript $n$ to denote that the estimator will depend on the value of $n$, with higher accuracy likely to be obtained for larger values of $n$.  We note that also because the likelihood is stochastic a different answer will be obtained for fixed $n$ if a different random seed is applied.  Since the optimisation in \eqref{eq:max_sl} is stochastic, \citet{Wood2010} applied a Markov chain Monte Carlo (MCMC) to explore the space of $\theta$ and select the value of $\theta$ that produced the highest value of the SL.  Some recent applications of the SL method have appeared in \citet{Hartig2014}, who used the FORMIND model for explaining complicated biological processes that occur in natural forests, and \citet{Brown2014}, who considered models for the transmission dynamics of avian influenza viruses in different bird types.

The synthetic likelihood approach has a strong connection with indirect inference, which is a classical method for obtaining point estimates of parameters of models with intractable likelihoods.  In the simulated quasi-maximum likelihood (SQML) approach of \citet{Smith1993}, an auxiliary model with a tractable likelihood function, $p_A(y|\phi)$, where $\phi$ is the parameter of that model, is used.   Define the function $\phi(\theta)$ as the relationship between the parameter of the model of interest and the auxiliary model.  This is often referred to as the binding function in the indirect inference literature.  The SQML method aims to maximise the auxiliary likelihood rather than the intractable likelihood of the model of interest
\begin{align*}
\hat{\theta} &= \max_{\theta}p_A(y_{obs}|\phi(\theta)).
\end{align*}   
Unfortunately the binding function is typically unavailable.  However, it can be estimated by generating $n$ iid datasets, $y_1,\ldots,y_n$, from the model of interest (the generative model) conditional on a value of $\theta$.  Define the auxiliary parameter estimate based on the $i$th simulated dataset as
\begin{align*}
\phi_{y_i} &= \arg \max_{\phi}p_A(y_i|\phi).
\end{align*}
Then we have
\begin{align*}
\phi(\theta) \approx \phi_n(\theta) = \frac{1}{n}\sum_{i=1}^n \phi_{y_i}.
\end{align*}
The binding function is defined as $\phi_n(\theta) \rightarrow \phi(\theta)$ as $n \rightarrow \infty$.  The SQML estimator then becomes
\begin{align*}
\hat{\theta}_n &= \max_{\theta}p_A(y_{obs}|\phi_n(\theta)).
\end{align*}
The synthetic likelihood falls within the SQML framework but where $y_{obs}$ has been reduced to $s_{obs}$ and the density of the multivariate normal distribution is used for $p_A$.

\subsection{Bayesian synthetic Likelihood}

An intuitive approach to incorporating SL into a Bayesian framework involves combining the prior $\pi(\theta)$ with the synthetic likelihood, which induces the following approximate posterior
\begin{align*}
\pi_n(\theta|s_{obs}) &\propto  {\cal N}(s_{obs};\mu_n(\theta),\Sigma_n(\theta))\pi(\theta),
\end{align*}
where the subscript $n$ denotes that the approximate posterior depends on the choice of $n$.  \citet{DrovandiPettittLee2015} consider a general framework called parametric Bayesian indirect likelihood (pBIL), where the likelihood of some auxiliary model with parameter $\phi$, $p_A(y_{obs}|\phi(\theta))$, is used to replace the intractable likelihood of the actual or generative model, $p(y_{obs}|\theta)$.  Since the binding function is generally not available in closed form, it can be estimated by simulation via drawing $n$  iid datasets from the generative model and fitting the auxiliary model to this simulated data (as in the SQML method mentioned previously), producing $\phi_n(\theta)$.  \citet{DrovandiPettittLee2015} demonstrate that the resulting approximate posterior depends on $n$, since in general $p_A(y_{obs}|\phi_n(\theta))$ is not an unbiased estimate of $p_A(y_{obs}|\phi(\theta))$ even when $\phi_n(\theta)$ is an unbiased estimate of $\phi(\theta)$.  We note that when non-negative and unbiased likelihood estimates are used within Monte Carlo methods such as MCMC \citep{Andrieu2009} and sequential Monte Carlo (SMC, \citet{Chopin2012}) algorithms, the resulting target distribution is the posterior based on the originally intended likelihood function.  Such approaches are referred to as pseudo-marginal or exact-approximate methods in the literature.  BSL fits within the pBIL framework, but where the auxiliary model is applied at a summary statistic level rather than the full data level and that the auxiliary model is the multivariate normal distribution, so that the auxiliary parameter estimates have an analytic expression as shown in \eqref{eq:mean_cov_estimates}.  Despite the fact that we use unbiased estimators for $\mu(\theta)$ and $\Sigma(\theta)$ (under the normality assumption) it is clear that ${\cal N}(s_{obs};\mu_n(\theta),\Sigma_n(\theta))$ is not an unbiased estimate of ${\cal N}(s_{obs};\mu(\theta),\Sigma(\theta))$.  Therefore the BSL posterior is inherently dependent on $n$.  However, under the assumption that the model is able to recover the observed statistic, \citet{Price2016} present extensive empirical evidence that the BSL posterior is remarkably insensitive to $n$.  Further, some empirical evidence demonstrates that BSL shows some robustness to the lack of multivariate normality.

\citet{Price2016} developed a new BSL method that uses an exactly unbiased estimator of the normal likelihood, which is developed by \citet{Ghurye1969}.     Using the notation of \citet{Ghurye1969}, let
\begin{align*}
c(k,v) &= \frac{2^{-kv/2}\pi^{-k(k-1)/4}}{\prod_{i=1}^k\Gamma \left(\frac{1}{2}(v-i+1)\right)},
\end{align*}
and for a square matrix $A$ write $\psi(A) = |A|$ if $A>0$ and $\psi(A) = 0$ otherwise, where $|A|$ is the determinant of $A$ and $A > 0$ means that $A$ is positive definite.  The result of \citet{Ghurye1969} shows that an exactly unbiased estimator of ${\cal N}(s_{obs};\mu(\theta),\Sigma(\theta))$ is (in the case where the summary statistics are normal and $n > d+3$)
\begin{align*}
\hat{p}_A(s_{obs}|\phi(\theta)) &= (2\pi)^{-d/2}\frac{c(d,n-2)}{c(d,n-1)(1-1/n)^{d/2}}|M_n(\theta)|^{-(n-d-2)/2}  \\ 
& \qquad \psi \left(  M_n(\theta) - (s_{y} - \mu_n(\theta))(s_{y} - \mu_n(\theta))^\top/(1-1/n)   \right)^{(n-d-3)/2},
\end{align*}
where $M_n(\theta) = (n-1)\Sigma_n(\theta)$.  It is interesting to note that this estimator is a mixture of a discrete and a continuous random variable (a realisation of the estimator can be identically 0 with positive probability).  Thus, if this estimator is used within a Monte Carlo method, the target distribution is proportional to ${\cal N}(s_{obs};\mu(\theta),\Sigma(\theta))\pi(\theta)$ regardless of the value of $n$ (under the multivariate normality assumption).  \citet{Price2016} referred to this method as uBSL, where `u' denotes unbiased.

To sample from the BSL posteriors, an MCMC algorithm can be used, for example.  We refer to this as MCMC BSL, which is shown in Algorithm \ref{synth_alg}.  Given the insensitivity of the BSL posteriors to the value of $n$, it is of interest to maximise the computational efficiency of the MCMC method.  For large $n$, the SL is estimated with high precision but the cost per iteration is high.   Conversely, for small $n$, the cost per iteration is low but the SL is estimated less precisely, which reduces the MCMC acceptance rate.  \citet{Price2016} found empirically that the value of $n$ that leads to an estimated log SL (at a $\theta$ with high BSL posterior support) with a standard deviation of roughly 2 produces efficient results.  However, \citet{Price2016} also found that there a wide variety of $n$ values that lead to similar efficiency.  When the unbiased SL is used in place of the SL shown in Algorithm \ref{synth_alg}, the MCMC uBSL algorithm is obtained.  In the examples of \citet{Price2016}, MCMC BSL and MCMC uBSL have a similar efficiency.  We also note that the MCMC BSL posteriors appear to exhibit very slow convergence when starting at a point with negligible posterior support.  The reason for this is that the SL is estimated with a large variance when the observed statistic $s_{obs}$ lies in the tail of the actual SL.  Thus additional research is required on more sophisticated methods for sampling from the BSL posteriors.

\begin{algorithm}
	\caption{MCMC BSL algorithm.  The inputs required are the summary statistic of the data, $s_{obs}$, the prior distribution, $p(\theta)$, the proposal distribution $q$, the number of iterations, $T$, and the initial value of the chain $\theta^{0}$.  The output is an MCMC sample $(\theta^{0},\theta^{1}, \ldots, \theta^{T})$ from the BSL posterior.  Some samples can be discarded as burn-in if required.}
	\label{synth_alg}
	\begin{algorithmic}[1]
		\STATE Simulate $s_{1:n} \stackrel{\mathrm{iid}}{\sim} p(\cdot|\theta^{0})$
		\STATE Compute $\phi^{0}=(\mu_n(\theta^{0}),\Sigma_n(\theta^{0}))$ 
		\FOR{$i = 1$ to $T$}
		\STATE Draw $\theta^{*} \sim q(\cdot|\theta^{i-1})$
		\STATE Simulate $s_{1:n}^{*} \stackrel{\mathrm{iid}}{\sim} p(\cdot|\theta^{*})$ 
		\STATE Compute $\phi^{*}=(\mu_n(\theta^{*}),\Sigma_n(\theta^{*}))$ 
		\STATE Compute $r=\min\left(1,\frac{\mathcal{N}(s_{obs};\mu_n(\theta^*),\Sigma_n(\theta^*))p(\theta^{*})q(\theta^{i-1}|\theta^{*})}{\mathcal{N}(s_{obs};\mu_n(\theta^{i-1}),\Sigma_n(\theta^{i-1}))p(\theta^{i-1})q(\theta^{*}|\theta^{i-1})}\right)$
		\IF{$\mathcal{U}(0,1)<r$}
		\STATE Set $\theta^{i}=\theta^{*}$ and $\phi^{i}=\phi^{*}$
		\ELSE
		\STATE Set $\theta^{i}=\theta^{i-1}$ and $\phi^{i}=\phi^{i-1}$
		\ENDIF
		\ENDFOR
	\end{algorithmic}
\end{algorithm}

The BSL method has been applied in the literature.  \citet{Fasiolo2014} used BSL for posterior inference for state space models in ecology and epidemiology based on data reduction and compared it with particle Markov chain Monte Carlo \citep{Andrieu2010}.  \citet{Hartig2014} implemented BSL for a forest simulation model.

BSL could be seen as a direct competitor with ABC as they are both simulation-based methods and differ only in the way the intractable summary statistic likelihood is approximated.  Importantly, BSL does not require the user to select a discrepancy function, as one is naturally induced via the multivariate normal approximation.  The simulated summary statistics in BSL are automatically scaled, whereas an appropriate weighting matrix to compare summary statistics in ABC must be done manually.   As noted in \citet{Blum2009} and \citet{DrovandiPettittLee2015}, ABC uses a non-parametric approximation of the summary statistic likelihood based on similar procedures used in kernel density estimation.  From this point of view, the ABC approach may be more accurate when the summary statistic $s_{obs}$ is low dimensional, however the accuracy/efficiency trade-off is less clear when the summary statistic $s_{obs}$ is high dimensional.  \citet{Price2016} demonstrated on a toy example that BSL becomes increasingly more computationally efficient than ABC as the dimension of the summary statistic grows beyond 2.  Furthermore, \citet{Price2016} demonstrated that BSL outperformed ABC in a cell biology application with 145 summary statistics.

\subsection{Accelerating synthetic likelihood}

As with ABC, the SL method is very simulation intensive.  There have been several attempts in the literature to accelerate the SL method by reducing the number of model simulations required.  \citet{Meeds2014} assumed that the summary statistics are independent and during their MCMC BSL algorithm fit a Gaussian process (GP) to each summary statistic output as a function of the model parameter $\theta$.  The Gaussian process (GP) is then used to predict the model output at proposed values of $\theta$, provided that the prediction is accurate enough.  If the GP prediction cannot be performed with sufficient accuracy, more model simulations are taken at that proposed $\theta$ and the GP is re-fit for each summary statistic.  The independence assumption of the summary statistics is questionable, and may overstate the information contained in $s_{obs}$.  

In contrast, \citet{Wilkison2014} used a GP to model the SL as a function of $\theta$ directly and use the GP to predict the SL at new values of $\theta$.  The GP is fit using a history matching approach \citep{Craig1997}.  Once the final GP fit is obtained, an MCMC algorithm is used with the GP prediction used in place of the SL.

\citet{Moores2014} considered accelerating Bayesian inference for the Potts model, which is a complex single parameter spatial model.  Simulations are performed across a pre-defined grid with the mean and standard deviation of the summary statistic (which turns out to be sufficient in the case of the Potts model, as it belongs to the exponential family) estimated from these simulations.  Non-parametric regressions are then fitted individually to the mean and standard deviation estimates in order to produce an estimate of the mappings $\mu(\theta)$ and $\sigma(\theta)$ across the space of $\theta$, where $\sigma$ is the standard deviation of the summary statistic.  The regressions are then able to predict the mean and standard deviation of the summary statistic at $\theta$ values not contained in the grid.  Further, the regression also smooths out the mappings, which are estimated using a finite value of $n$.  The estimated mapping is then used in a sequential Monte Carlo Bayesian algorithm. 

\subsection{Example}

Cell motility, cell proliferation and cell-to-cell adhesion play an important role in collective cell spreading, which is critical to many key biological processes, including skin cancer growth and wound healing (e.g. \citet{Cai2007, Treloar2013}). 
The main function of many medical treatments is to influence the biology underpinning these processes \citep{Decaestecker2007}. In order to measure the efficacy of such treatments, it is important that estimates of the parameters 
governing these cell spreading processes can be obtained along with some characterisation of their uncertainty. Agent-based computational models are frequently used to interpret these cell biology 
processes since they can produce discrete image-based and movie-based information which is ideally suited to collaborative investigations involving applied mathematicians and experimental cell biologists. 
Unfortunately, the likelihood functions for these models are computationally intractable, so standard statistical inferential methods for these models are not applicable.

To deal with the intractable likelihood, several papers have adopted an ABC approach to estimate the parameters \citep{Johnston2014,Vo2015,VoDiameter2014}. One difficulty with these cell biology applications is that the observed data are typically available as sequences of images and therefore it is not trivial to reduce the dimension of the summary statistic to a suitable level for ABC while simultaneously retaining relevant information contained in the images.  For example, \citet{Johnston2014} considered data collected every 5 minutes for 12 hours but only analyse images at 3 time points.  \citet{Vo2015} reduced images initially down to a 15 dimensional summary statistic, but perform a further dimension reduction based on the approach of \citet{Fearnhead2012} to ensure there is one summary statistic per parameter.

Here we will re-analyse the data considered in \citet{Treloar2013} and \citet{Vo2015}.  The data consist of images of spatially expanding human malignant melanoma cell populations.  To initiate each experiment, either 20,000 or 30,000 cells are approximately evenly distributed within a circular barrier, located
at the centre of the well. Subsequently, the barriers are lifted and population-scale images are recorded at either 24 hours or 48 hours, independently.   Furthermore, there are two types of experiments conducted. The first uses a treatment in order to inhibit cells giving birth (cell proliferation) while the second does not use the treatment.  Each combination of initial cell density, experimental elapsed time and treatment is repeated 3 times, for a total of 24 images.  The reader is referred to \citet{Treloar2013} for more details on the experimental protocol.  For simplicity, we consider here the 3 images related to using 20,000 initial cells, 24 hours elapsed experimental time and no cell proliferation inhibitor.

In order to summarise an image, \citet{Vo2015} considered 6 sub-regions along a transect of each image.  The position of the cells in these regions is mapped to a square lattice. The number of cells in each sub-region is counted, together with the number of isolated cells. A cell is identified as isolated if all of its nearest neighbours (north, south, east, west) are unoccupied.  For each region, these summary statistics are then averaged over the three independent replicates.  We refer to these 12 summary statistics as $\{c_i\}_{i=1}^6$ and $\{p_i\}_{i=1}^6$, where $c_i$ and $p_i$ are the number of cells and the percentage of isolated cells (averaged over the 3 images) for region $i$, respectively.  \citet{Vo2015} also estimated the radius of the entire cell colony using image analysis.  Thus \citet{Vo2015} included three additional summary statistics, $(r_{(1)},r_{(2)},r_{(3)})$, which are the estimated and ordered radii for the three images.  For more details on how these summary statistics are obtained, the reader is referred to \citet{Vo2015}.  This creates a total of 15 summary statistics, which is computationally challenging to deal with in ABC.  As mentioned earlier, \citet{Vo2015} found it beneficial to apply the technique of \citet{Fearnhead2012}, which uses a regression to estimate the posterior means of the model parameters from the initial summaries, which are then used as summary statistics in a subsequent ABC analysis.  Here we attempt to see whether or not BSL is able to accommodate the 15 summary statistics, and compare the results with the ABC approach of \citet{Vo2015}.

\citet{Treloar2013} and \citet{Vo2015} considered a discretised time and space (two-dimensional lattice) stochastic model to explain the cell spreading process of melanoma cells.  For more details on this model, the reader is referred to \citet{Treloar2013} and \citet{Vo2015}.  The model contains three parameters: $P_m$ (probability that an isolated agent can move to a neighbouring lattice site in one time step), $P_p$ (probability that an agent will attempt to proliferate and deposit a daughter at a neighbouring lattice site within a time step) and $q$ (the strength of cell-to-cell adhesion, that is, cells sticking together).  These model parameters can then be related to biologically relevant parameters such as cell diffusivity and the cell proliferation rate.  Here we will report inferences in terms of the parameter $\theta = (P_m,P_p,q)$.

Here we consider a simulated dataset with $P_m = 0.1$, $P_p = 0.0012$ and $q = 0.2$ (same simulated data as analysed in \citet{Vo2015}) and the real data.  We ran BSL using Algorithm \ref{synth_alg} with independent $\mathcal{U}(0,1)$ prior distributions on each parameter.  We used a starting value and proposal distribution for the MCMC based on the results provided in \citet{Vo2015}, so we do not apply any burn-in.  We also applied the uBSL algorithm.  The BSL approaches were run with $n=32$, 48, 80 and 112 (the independent simulations were farmed out across 16 processors of a computer node).  To compare the efficiency of the different choices of $n$ we considered the effective sample size (ESS) for each parameter divided by the total number of model simulations performed multiplied by a large constant scalar to increase the magnitude of the numbers (we refer to this as the normalised ESS).  

\begin{figure}
	\centering
	\includegraphics[height=0.3\textheight,width=1.1\textwidth]{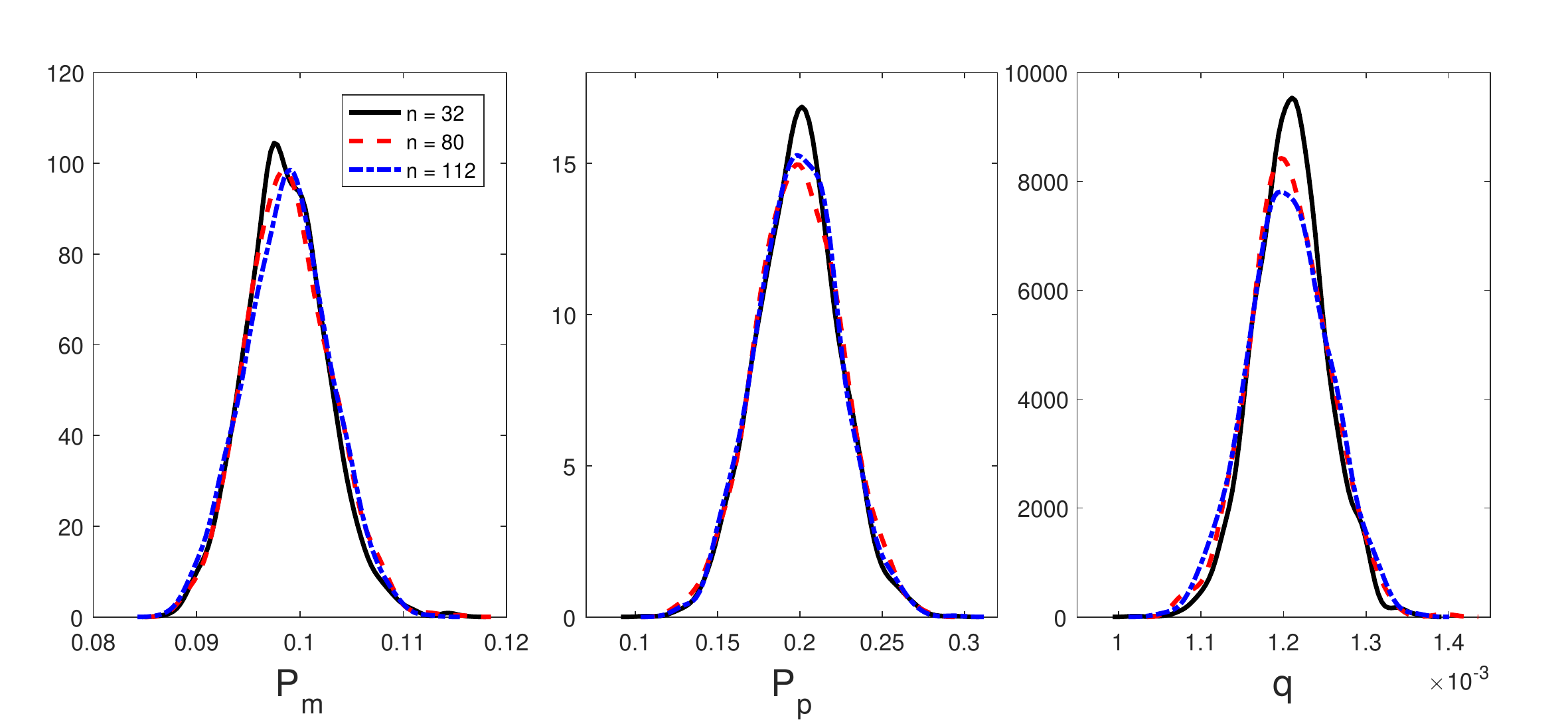}
	\caption{Posterior estimates for $P_m$, $P_p$ and $q$ based on the simulated data for the melanoma cell biology application using MCMC BSL for different values of $n$.}
	\label{fig:melanoma_sim_sl}
\end{figure}

\begin{figure}
	\centering
	\includegraphics[height=0.3\textheight,width=1.1\textwidth]{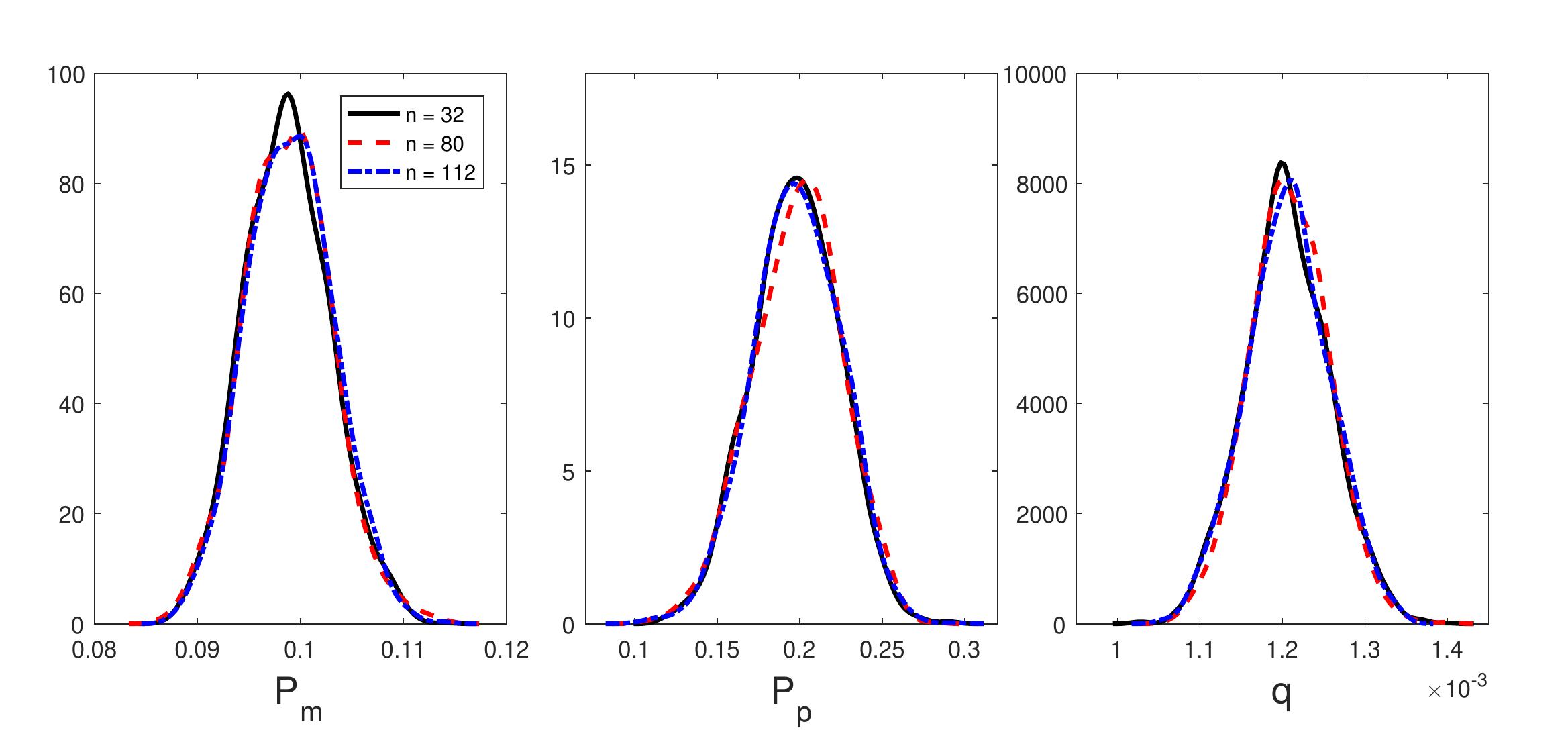}
	\caption{Posterior estimates for $P_m$, $P_p$ and $q$ based on the simulated data for the melanoma cell biology application using MCMC uBSL for different values of $n$.}
	\label{fig:melanoma_sim_usl}
\end{figure}

Marginal posterior distributions for the parameters obtained from BSL and uBSL for different values of $n$ are shown in Figures \ref{fig:melanoma_sim_sl} and \ref{fig:melanoma_sim_usl}, respectively.  It is evident that the posteriors are largely insensitive to $n$, which is consistent with the empirical results obtained in \citet{Price2016}.  The normalised ESS values and MCMC acceptance rates for the BSL approaches are shown in Table \ref{tab:cell_simulated} for different values of $n$.  The efficiency of BSL and uBSL appears to be similar.  The optimal value of $n$ out of the trialled values appears to be 32 or 48.  However, even $n=80$ is relatively efficient.  For $n=112$ the increase in acceptance rate is relatively small given the extra amount of computation required per iteration.

\begin{figure}
	\centering
	\includegraphics[height=0.3\textheight,width=1.1\textwidth]{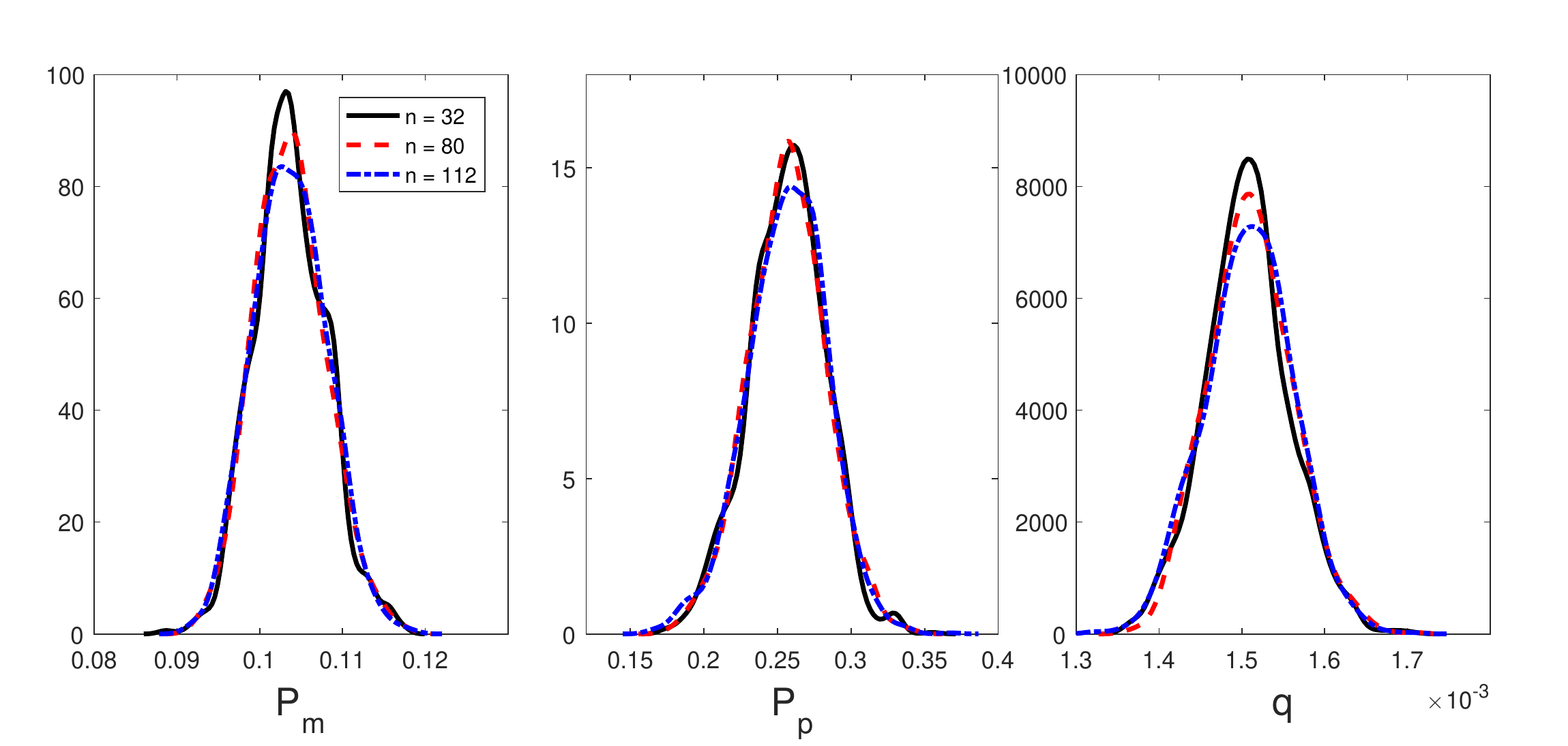}
	\caption{Posterior estimates for $P_m$, $P_p$ and $q$ based on the real data for the melanoma cell biology application using MCMC BSL for different values of $n$.}
	\label{fig:melanoma_real_sl}
\end{figure}

\begin{figure}
	\centering
	\includegraphics[height=0.3\textheight,width=1.1\textwidth]{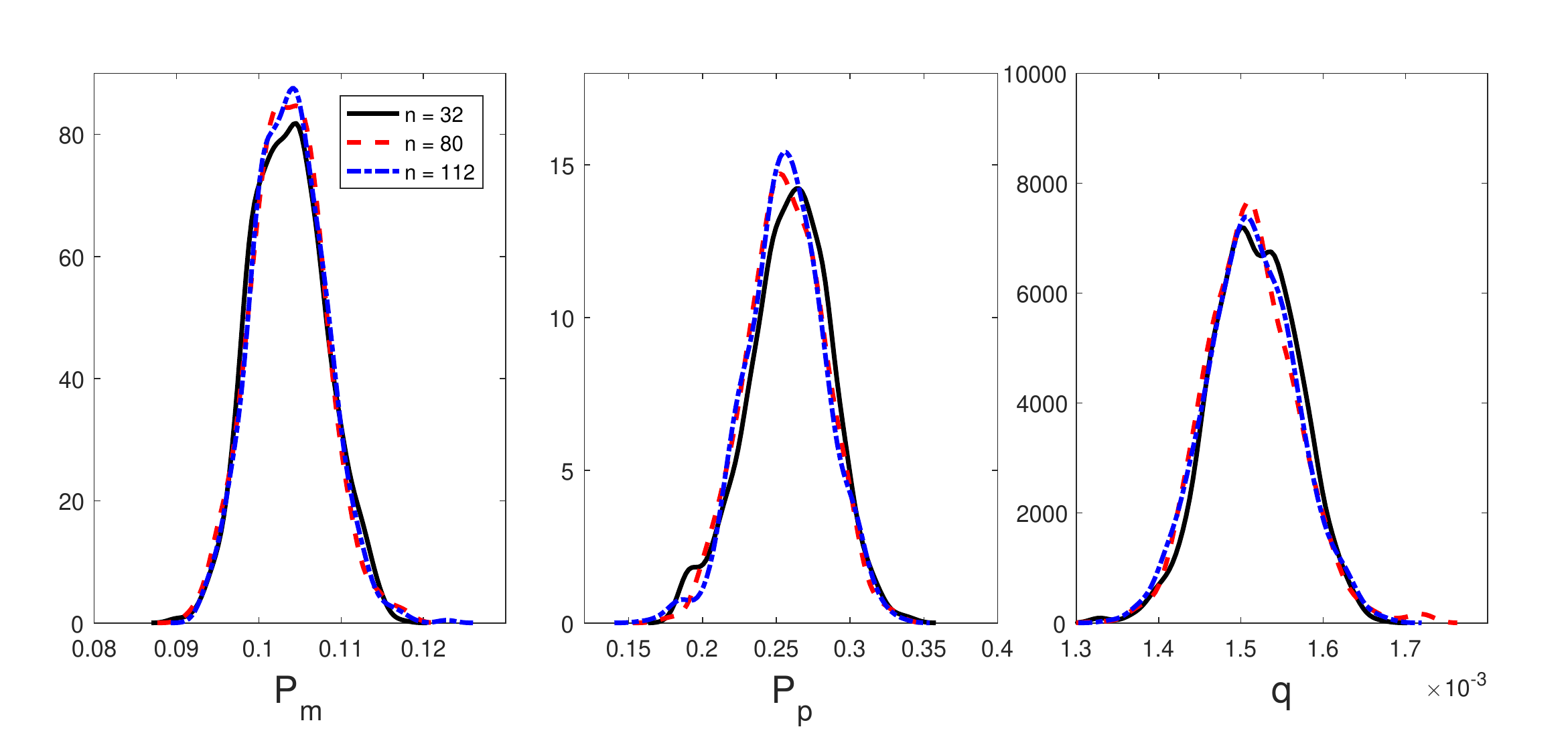}
	\caption{Posterior estimates for $P_m$, $P_p$ and $q$ based on the simulated data for the melanoma cell biology application using MCMC uBSL for different values of $n$.}
	\label{fig:melanoma_real_usl}
\end{figure}

We also applied the BSL approaches with similar settings to the real data.  The posterior results are presented in Figures \ref{fig:melanoma_real_sl} and \ref{fig:melanoma_real_usl}.  Again we found the results are relatively insensitive to $n$.  Table \ref{tab:cell_real} suggests that $n=48$ or $n=80$ are the most efficient choices for $n$ out of those trialled.  However, it is again apparent that there are a wide variety of $n$ values that lead to similar efficiency.

\begin{table}[!htp]
	\caption{Sensitivity of BSL/uBSL to $n$ for the simulated data of the cell biology example with regards to MCMC acceptance rate, normalised ESS for each parameter.}
	\centering
	\begin{tabular}{ccccc}
		\hline
		\hline
		$n$ & acc.\ rate (\%) & ESS $P_m$ & ESS $P_p$ & ESS $q$   \\
		\hline
		32  & 17/17   & 96/114  & 86/113 & 115/126    \\ 
		48  & 27/32 & 95/103  & 93/92  & 110/115   \\ 
		80  & 35/37   & 82/76  & 74/67 & 106/89     \\ 
		112  & 38/40   &  61/65 & 61/58 & 68/70     \\
		\hline
	\end{tabular}
	\label{tab:cell_simulated}
\end{table}   

\begin{table}[!htp]
	\caption{Sensitivity of BSL/uBSL to $n$ for the real data of the cell biology example with regards to MCMC acceptance rate, normalised ESS for each parameter.}
	\centering
	\begin{tabular}{ccccc}
		\hline
		\hline
		$n$ & acc.\ rate (\%) & ESS $P_m$ & ESS $P_p$ & ESS $q$   \\
		\hline
		32  & 8/9   & 46/51  &  38/45 &  41/43    \\ 
		48  & 17/18 & 76/71  & 56/63  & 70/54    \\ 
		80  & 27/28   &  66/64  &  66/60 & 68/58     \\ 
		112  & 32/33   &  58/60 & 51/54 & 54/43     \\
		\hline
	\end{tabular}
	\label{tab:cell_real}
\end{table} 


\begin{figure}
	\centering
	\includegraphics[height=0.3\textheight,width=1.1\textwidth]{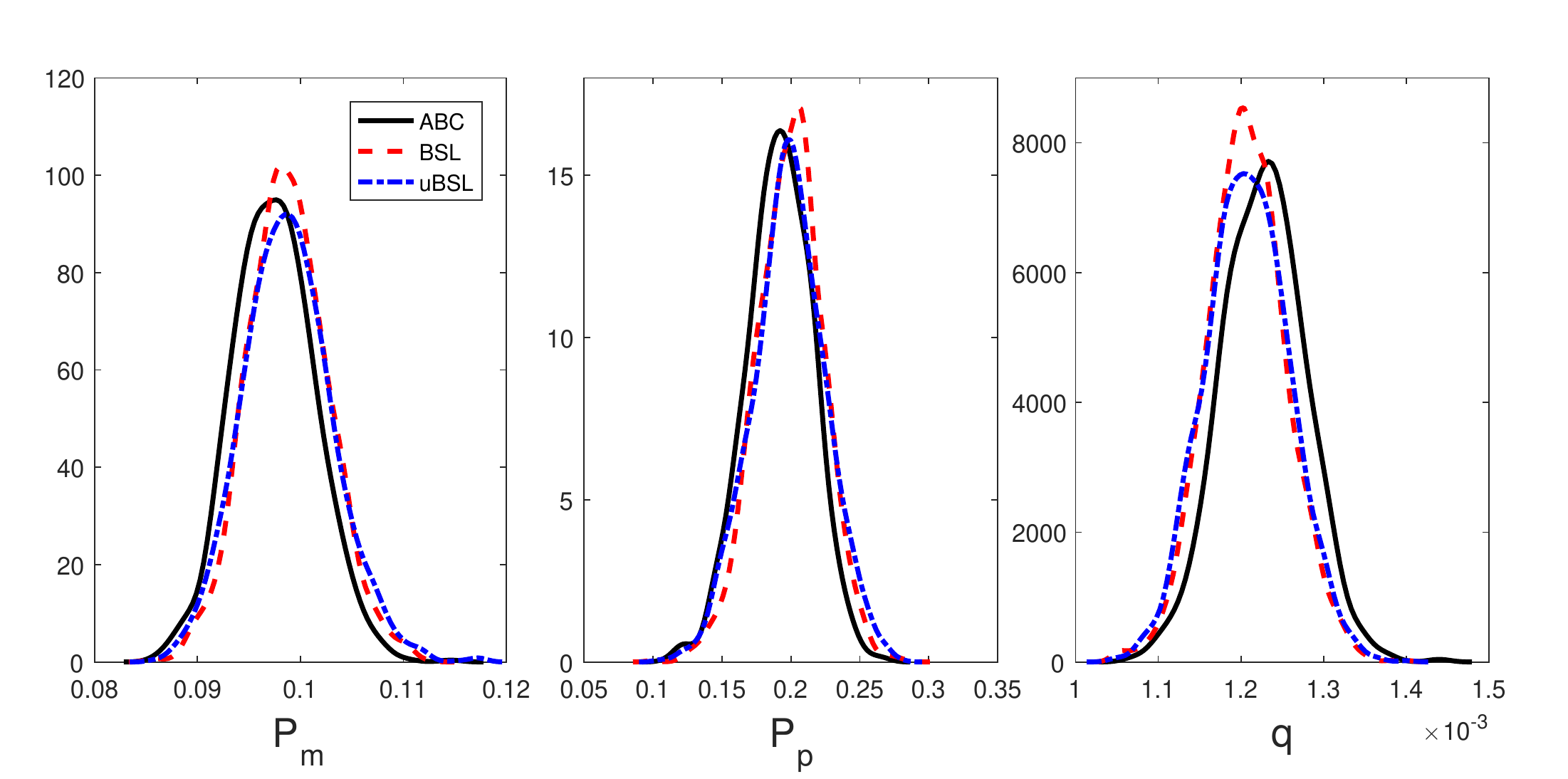}
	\caption{Posterior estimates for $P_m$, $P_p$ and $q$ for the melanoma cell biology application using the ABC approach of \citet{Vo2015} (solid), BSL (dash) and uBSL (dot-dash) based on simulated data with $P_m = 0.1$, $P_p = 0.0012$ and $q = 0.2$.  The BSL results are based on $n=48$.}
	\label{fig:melanoma_sim}
\end{figure} 


\begin{figure}
	\centering
	\includegraphics[height=0.3\textheight,width=1.1\textwidth]{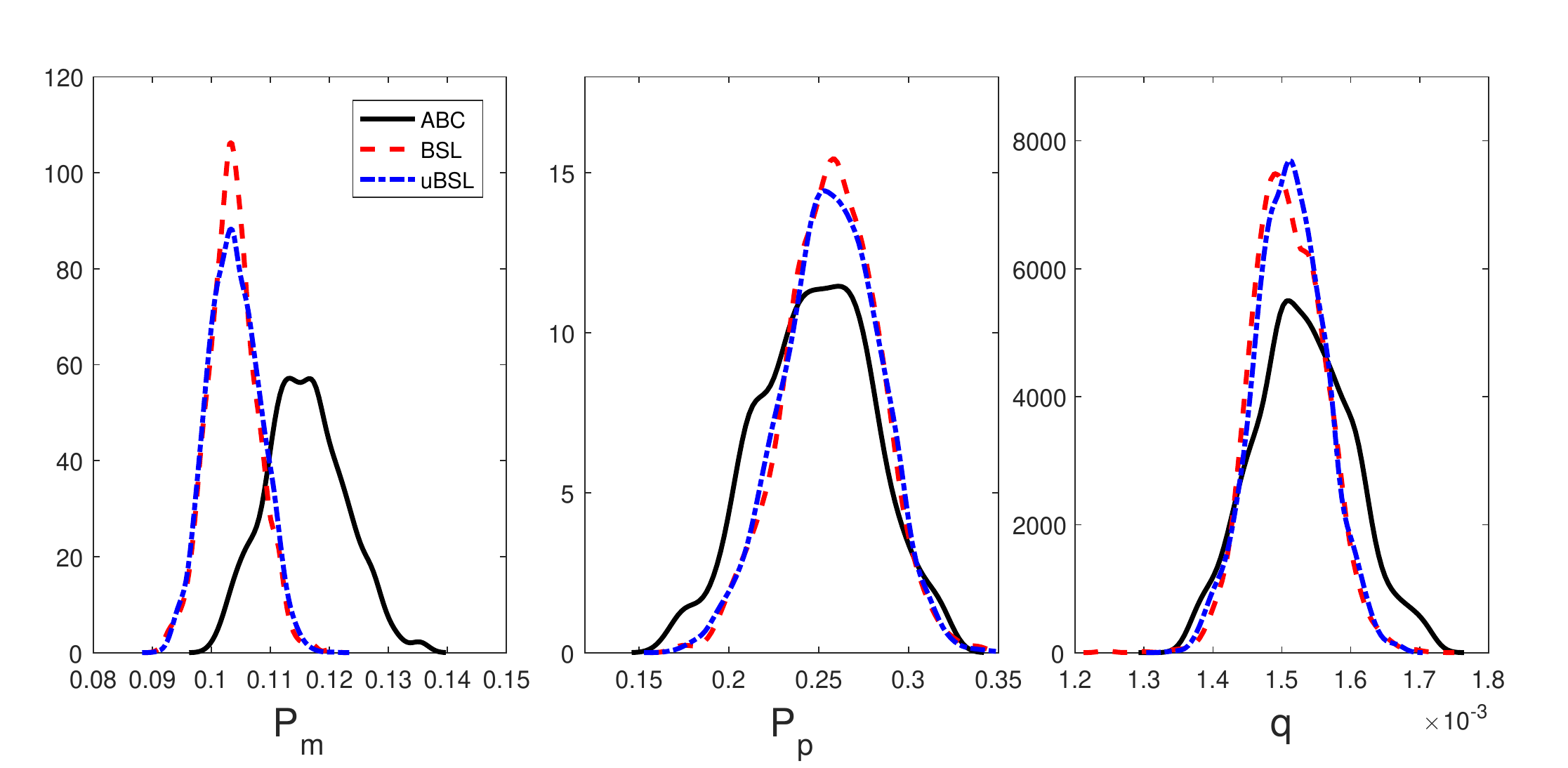}
	\caption{Posterior estimates for $P_m$, $P_p$ and $q$ for the melanoma cell biology application using the ABC approach of \citet{Vo2015} (solid), BSL (dash) and uBSL (dot-dash) based on real data.  The BSL results are based on $n=48$.}
	\label{fig:melanoma}
\end{figure}

\begin{figure}
	\centering
	\subfigure[$r_{(1)}$]{\includegraphics[height=0.15\textheight,width=0.3\textwidth]{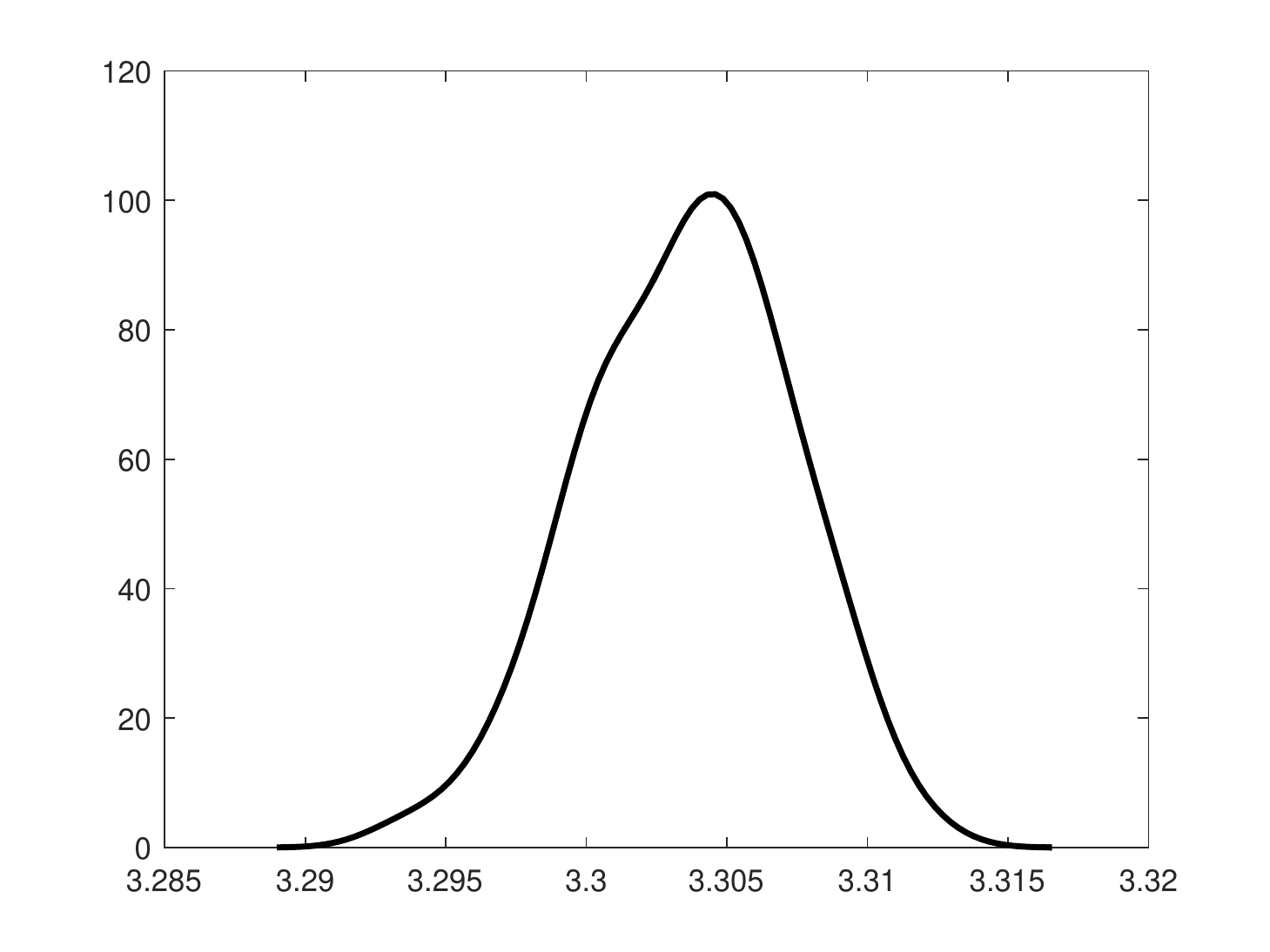}\label{figsub:ss1}}
	\subfigure[$r_{(2)}$]{\includegraphics[height=0.15\textheight,width=0.3\textwidth]{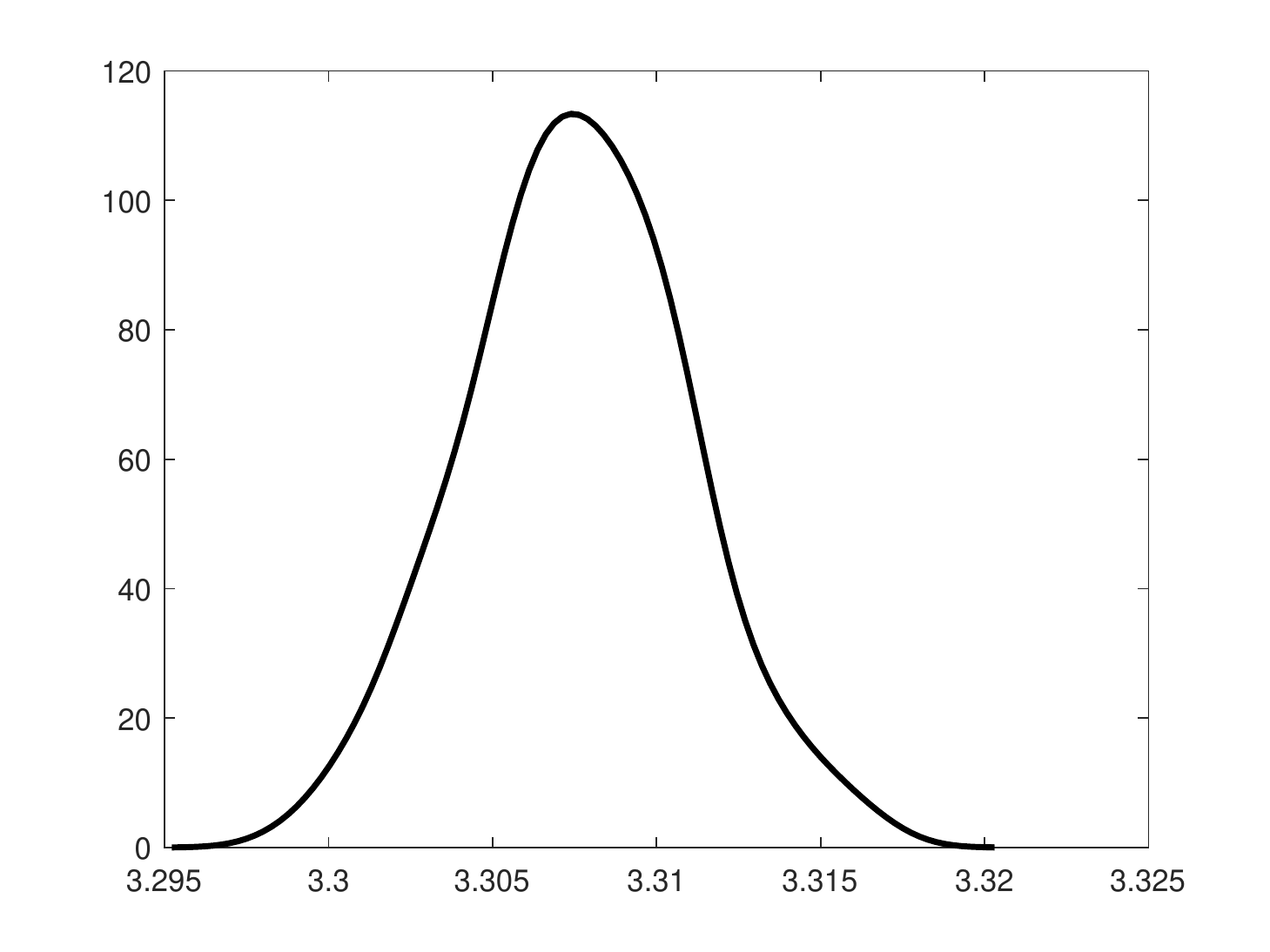}\label{figsub:ss2}}
	\subfigure[$r_{(3)}$]{\includegraphics[height=0.15\textheight,width=0.3\textwidth]{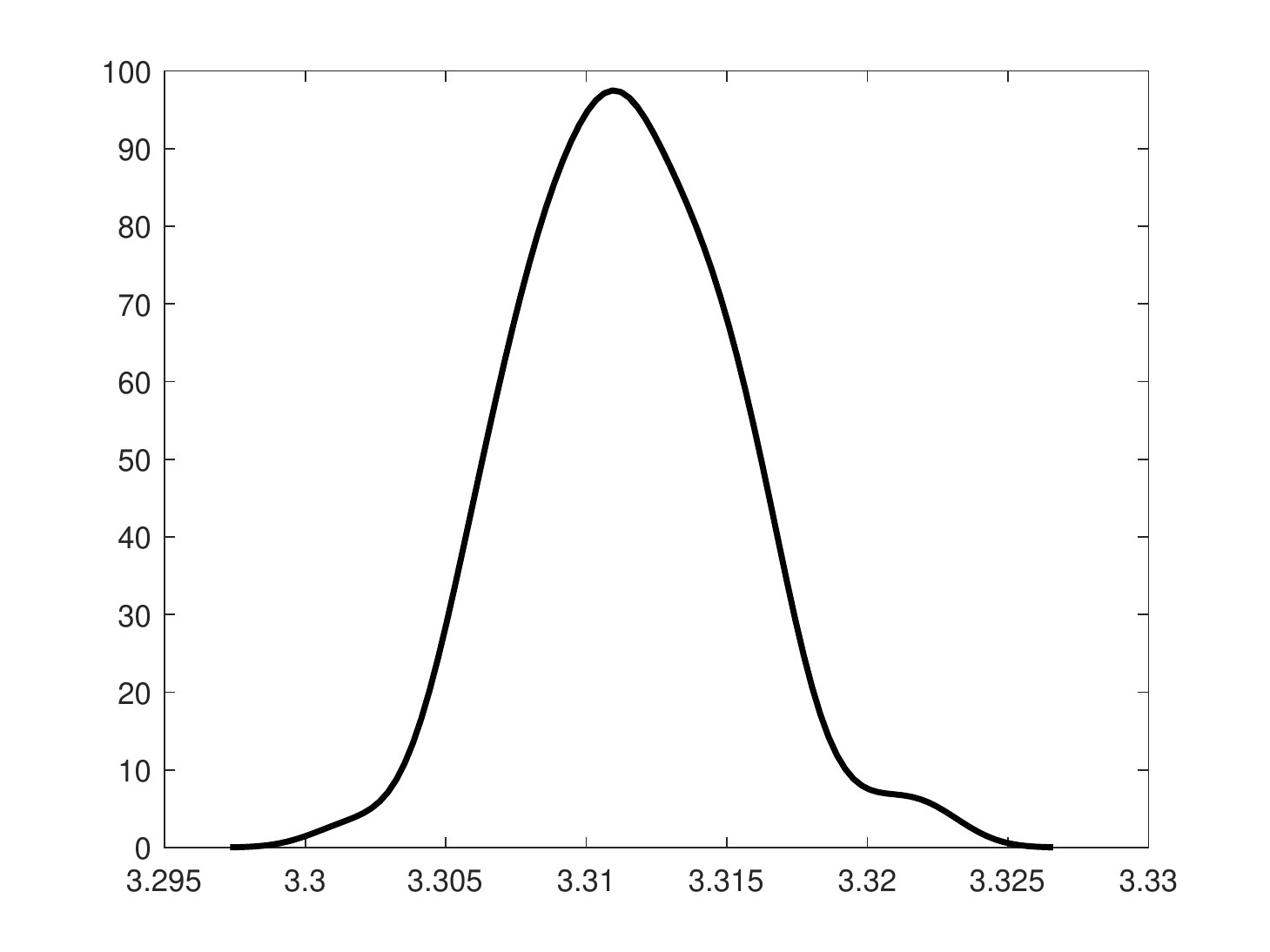}\label{figsub:ss3}}
	\subfigure[$p_1$]{\includegraphics[height=0.15\textheight,width=0.3\textwidth]{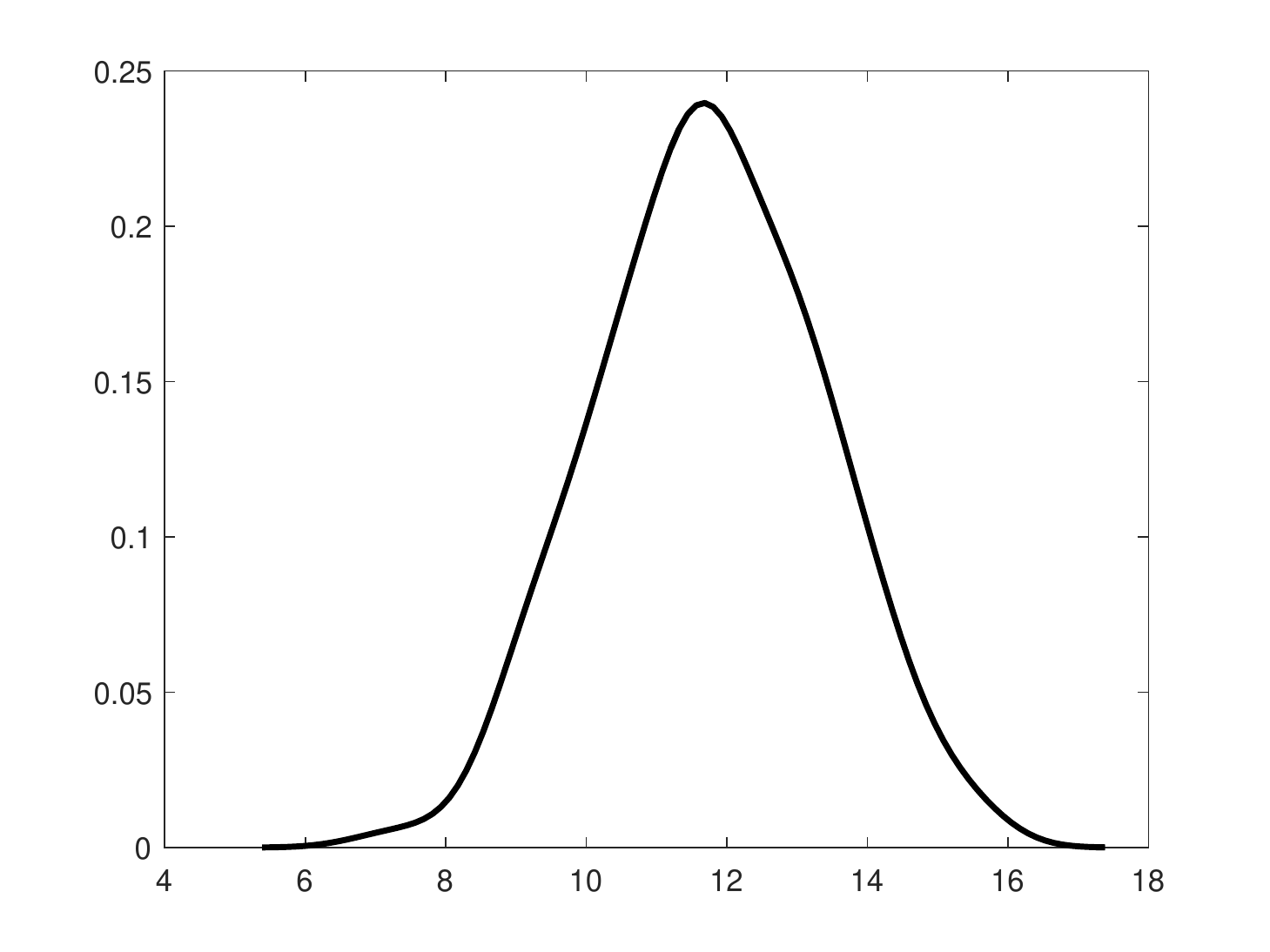}\label{figsub:ss4}}
	\subfigure[$p_2$]{\includegraphics[height=0.15\textheight,width=0.3\textwidth]{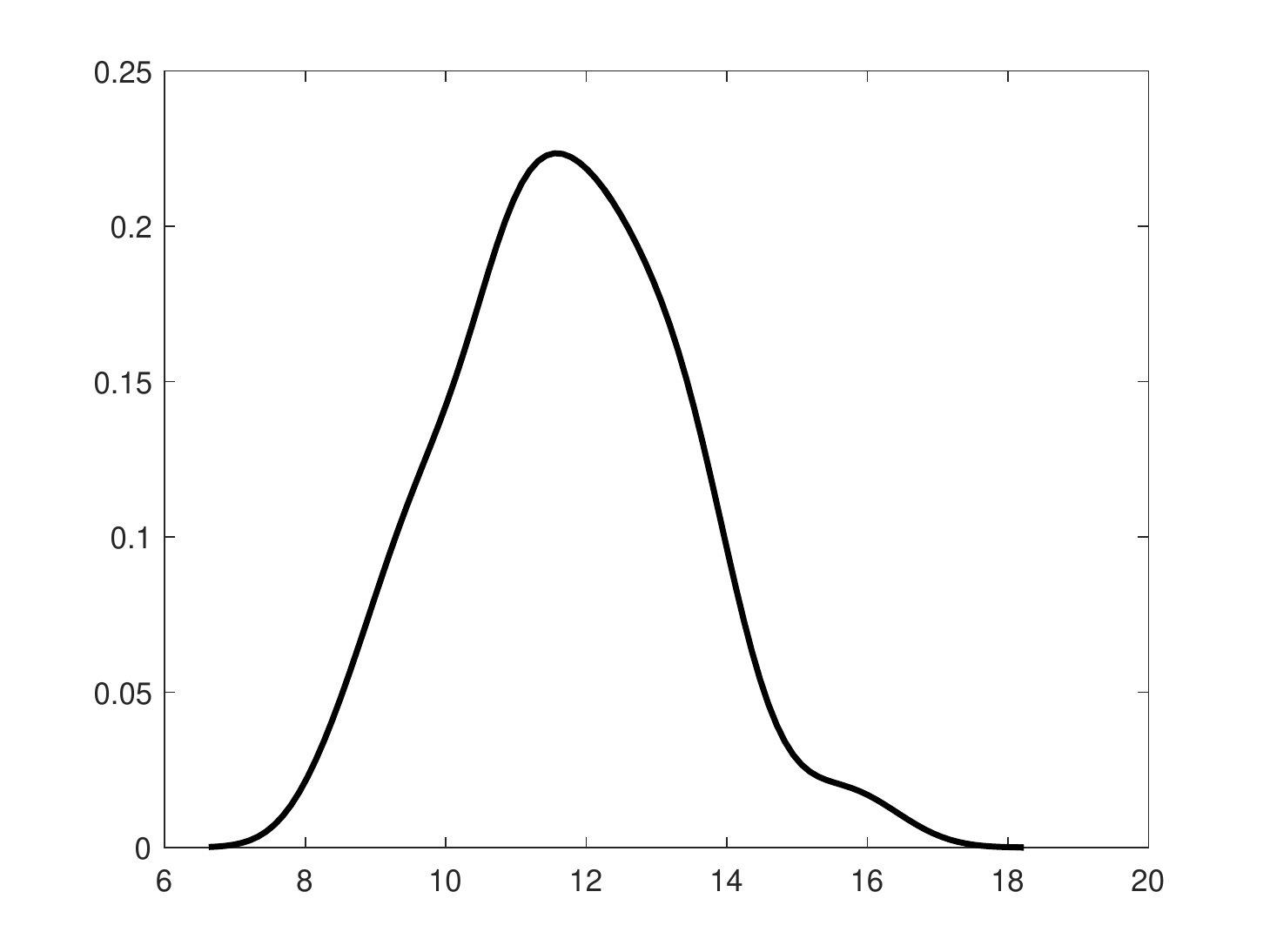}\label{figsub:ss5}}
	\subfigure[$p_3$]{\includegraphics[height=0.15\textheight,width=0.3\textwidth]{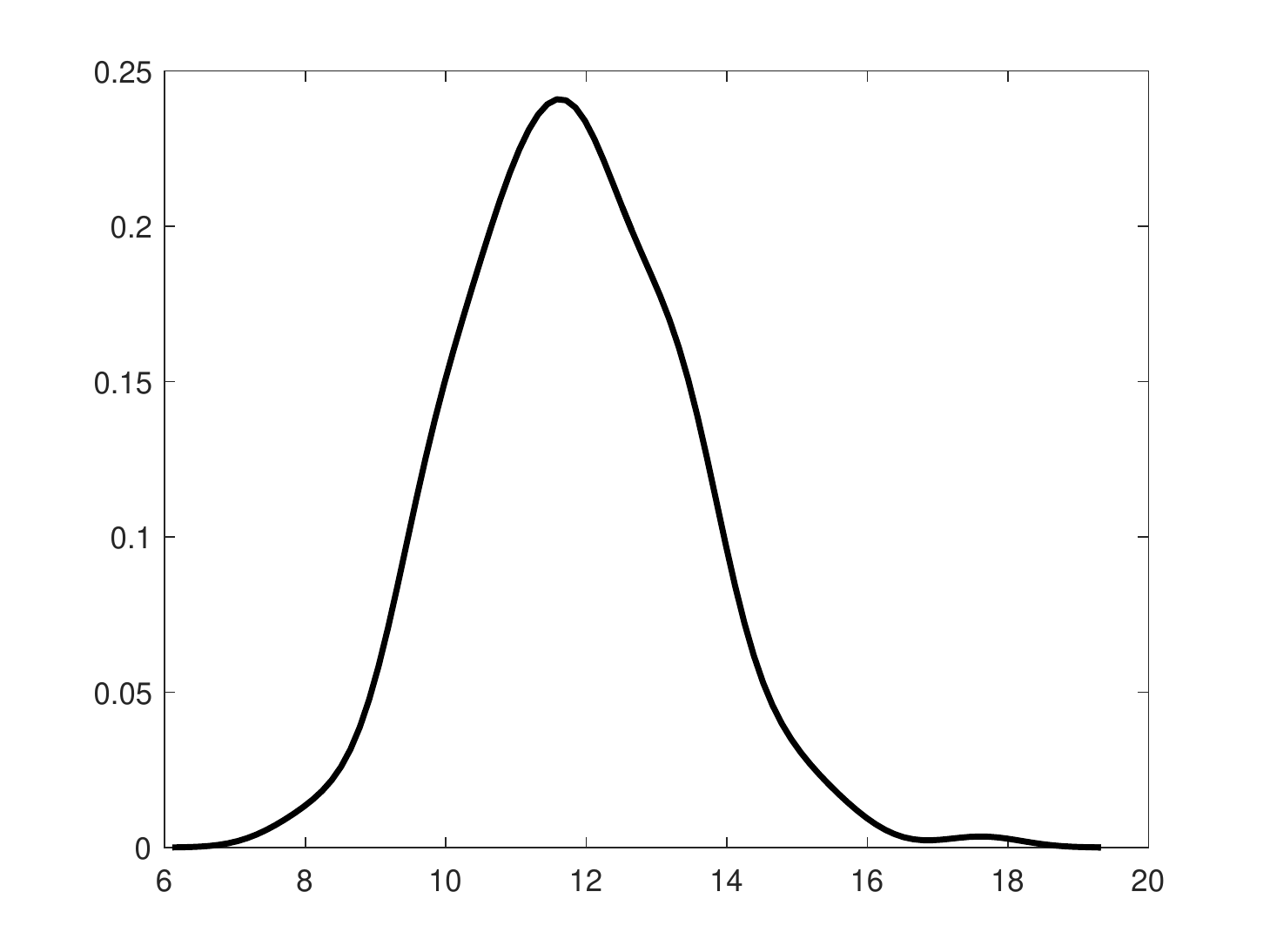}\label{figsub:ss6}}
	\subfigure[$p_4$]{\includegraphics[height=0.15\textheight,width=0.3\textwidth]{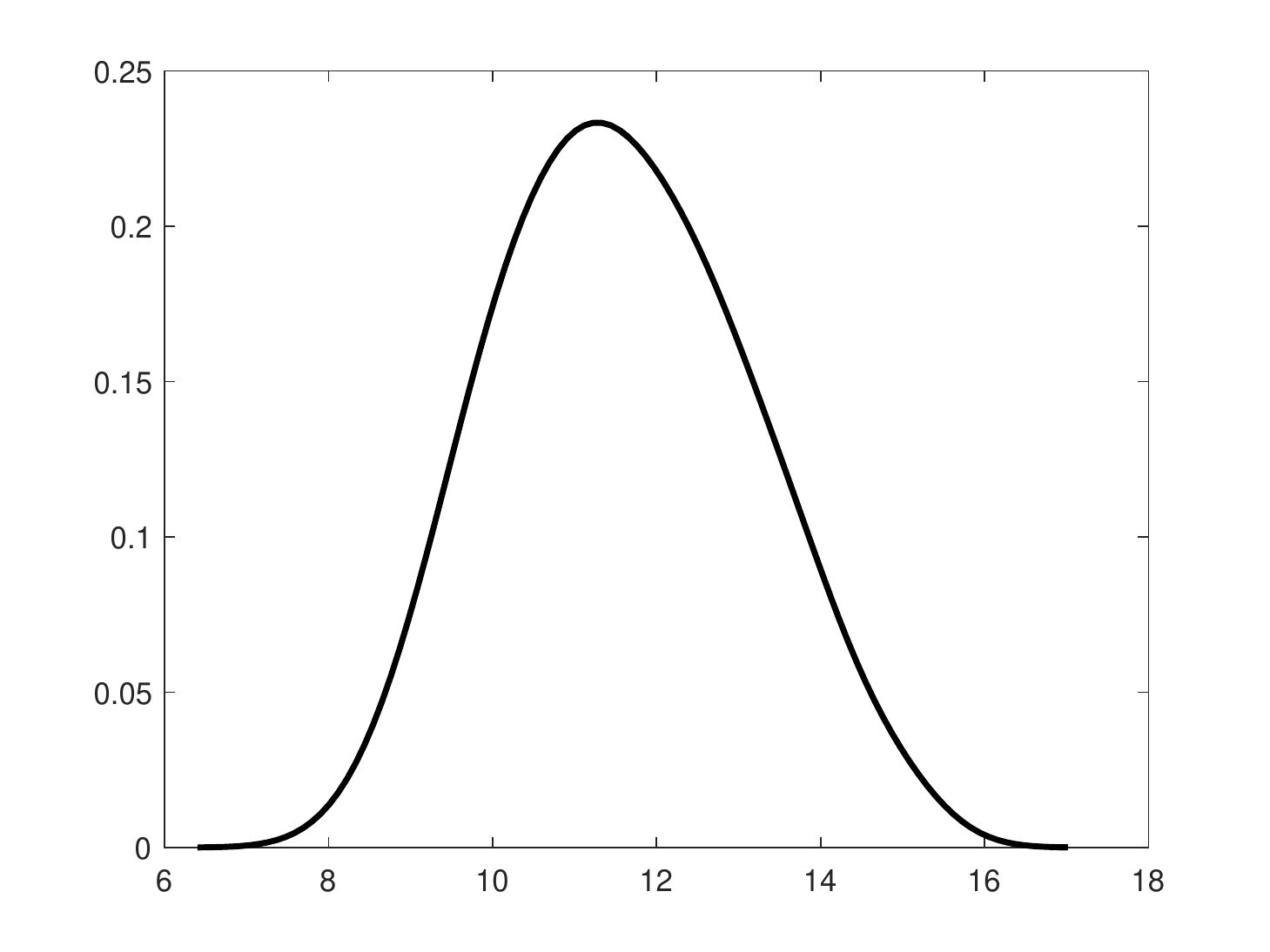}\label{figsub:ss7}}
	\subfigure[$p_5$]{\includegraphics[height=0.15\textheight,width=0.3\textwidth]{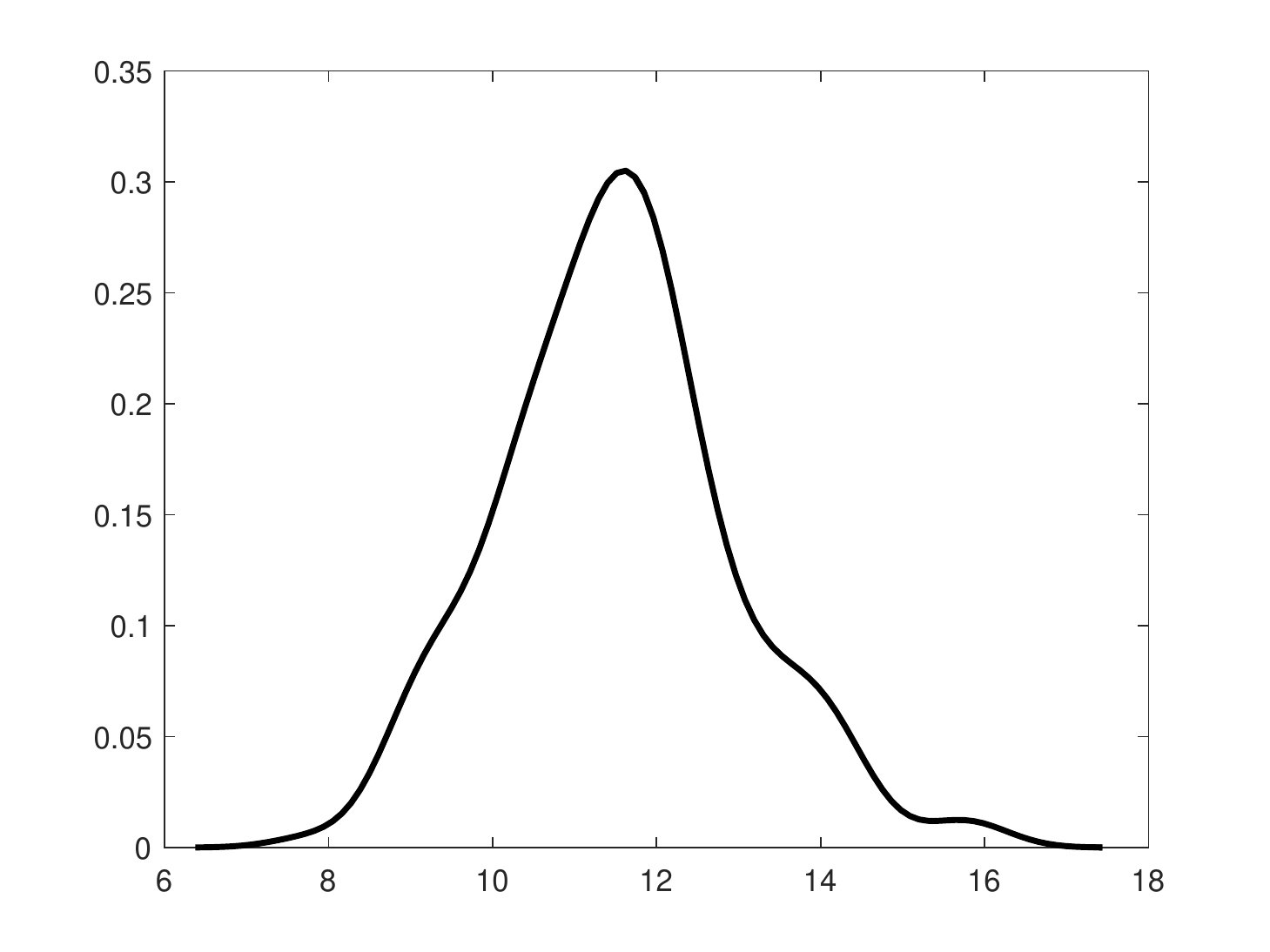}\label{figsub:ss8}}
	\subfigure[$p_6$]{\includegraphics[height=0.15\textheight,width=0.3\textwidth]{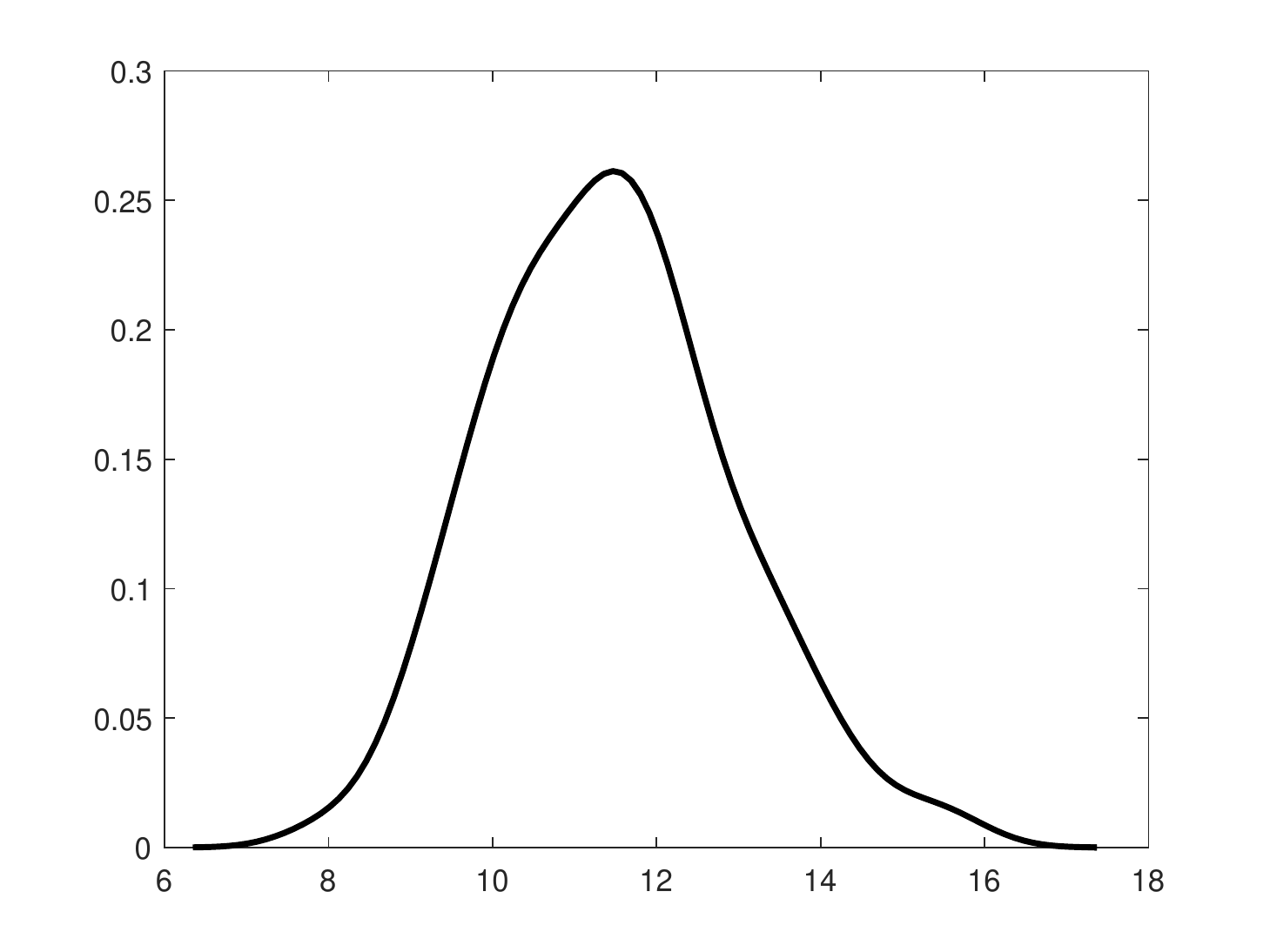}\label{figsub:ss9}}
	\subfigure[$c_1$]{\includegraphics[height=0.15\textheight,width=0.3\textwidth]{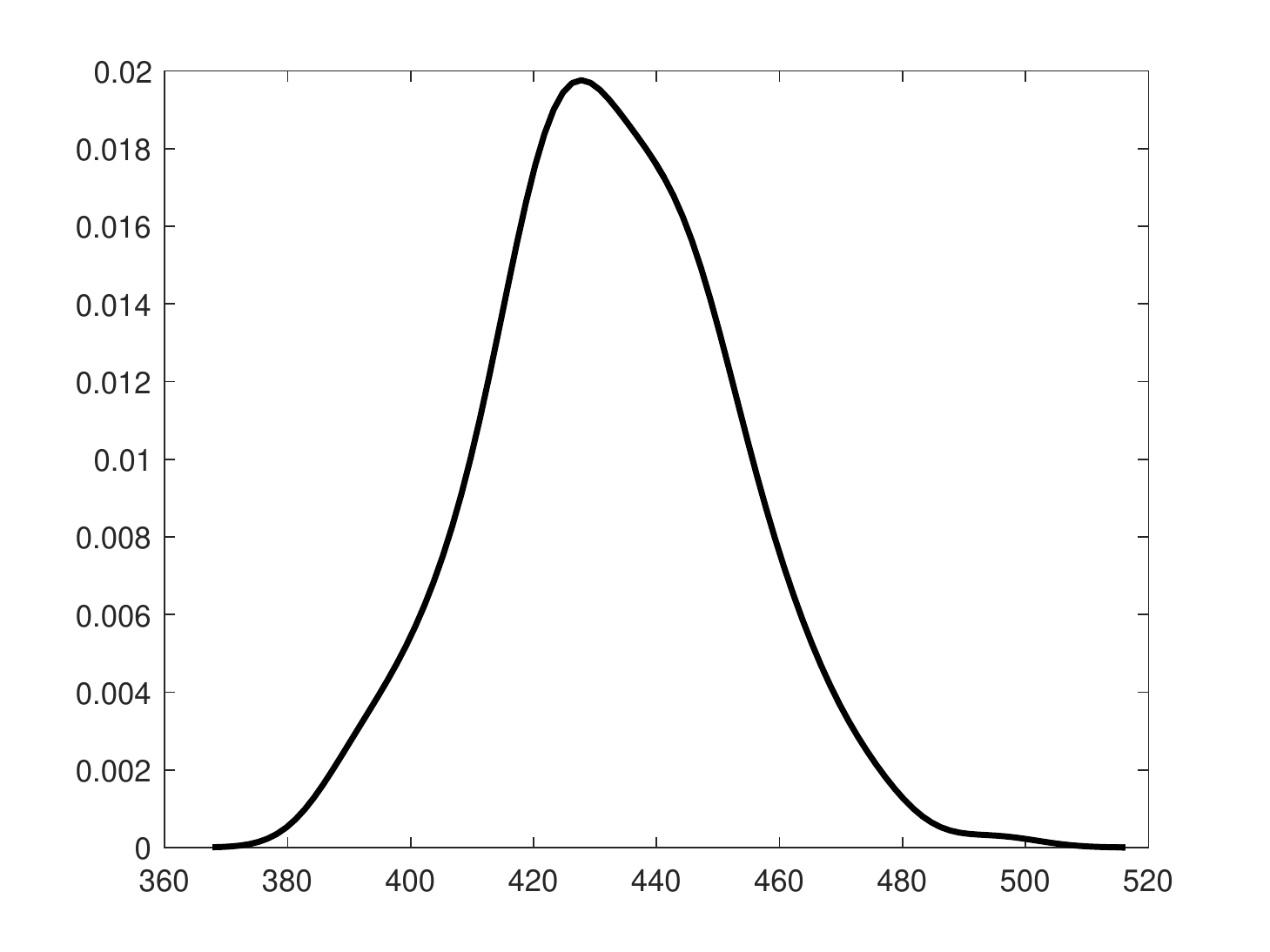}\label{figsub:ss10}}
	\subfigure[$c_2$]{\includegraphics[height=0.15\textheight,width=0.3\textwidth]{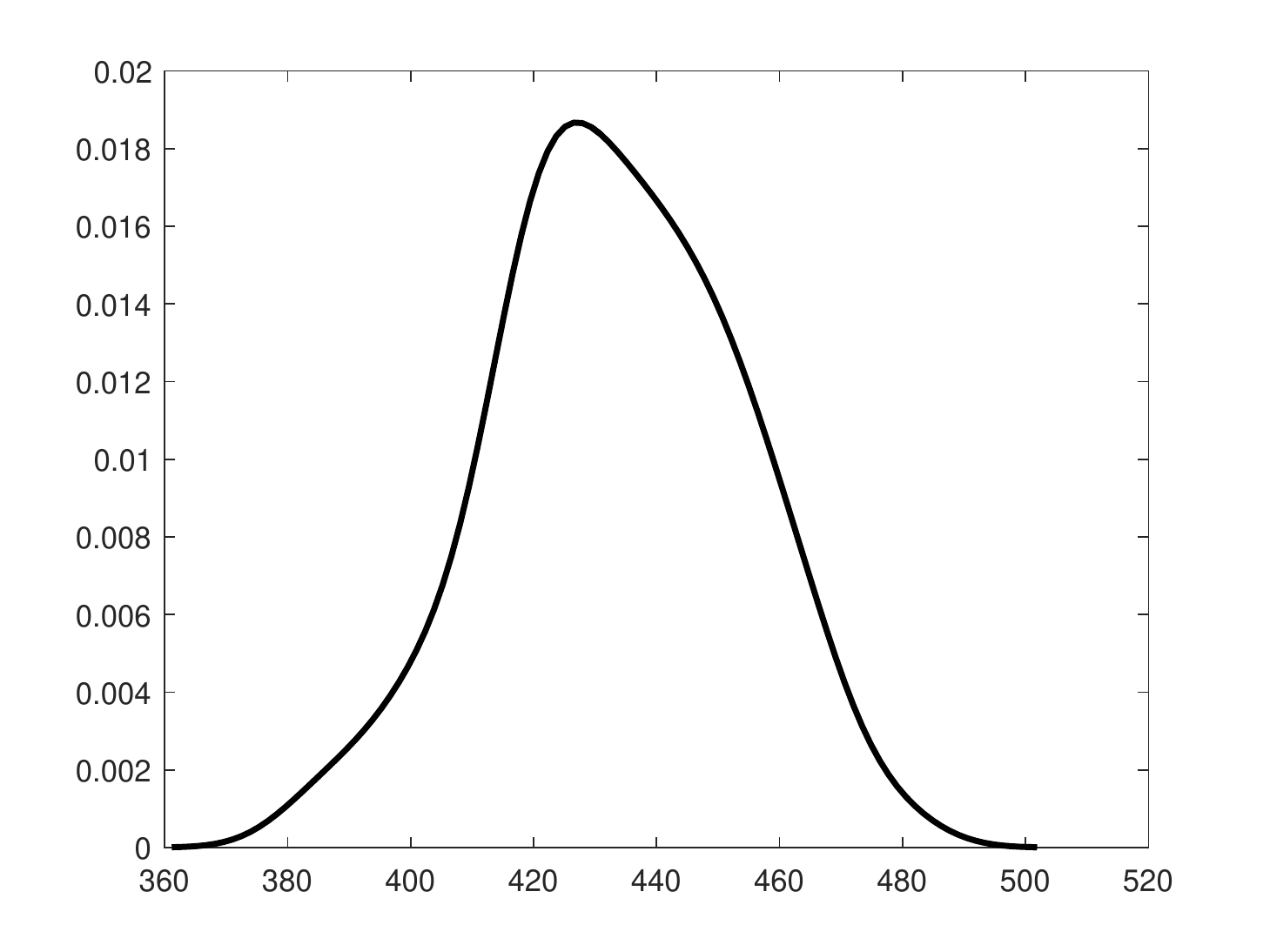}\label{figsub:ss11}}
	\subfigure[$c_3$]{\includegraphics[height=0.15\textheight,width=0.3\textwidth]{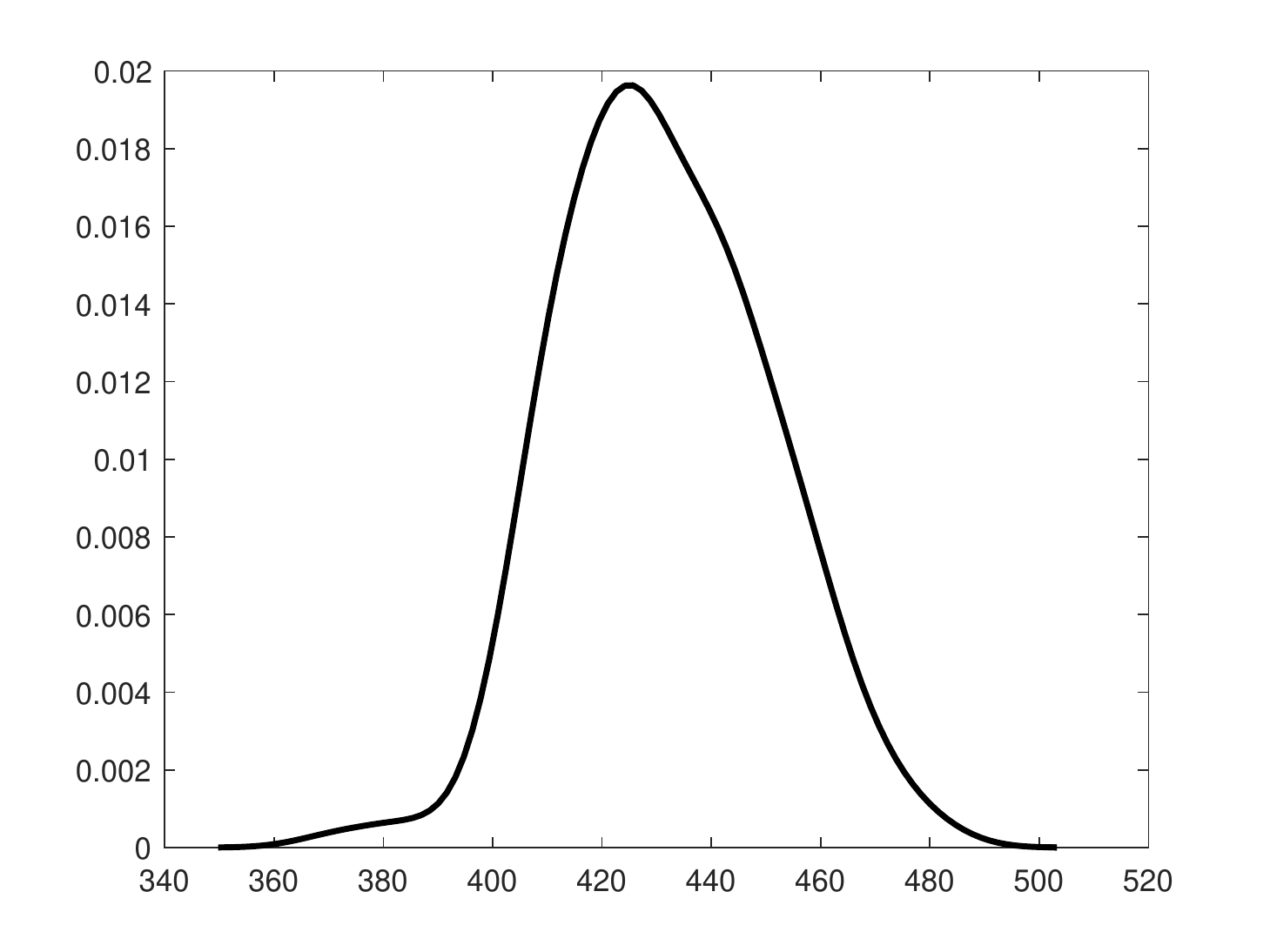}\label{figsub:ss12}}
	\subfigure[$c_4$]{\includegraphics[height=0.15\textheight,width=0.3\textwidth]{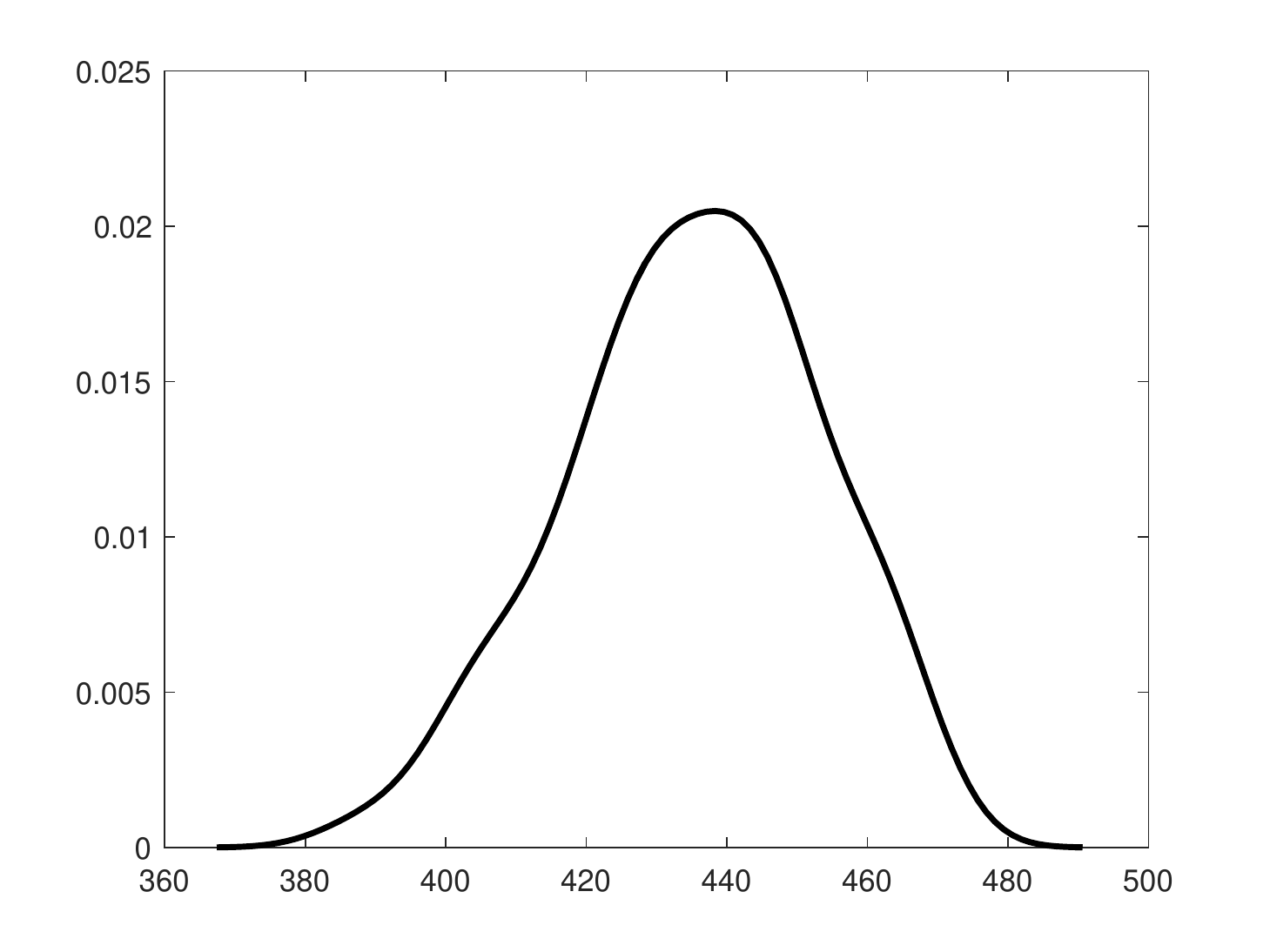}\label{figsub:ss13}}
	\subfigure[$c_5$]{\includegraphics[height=0.15\textheight,width=0.3\textwidth]{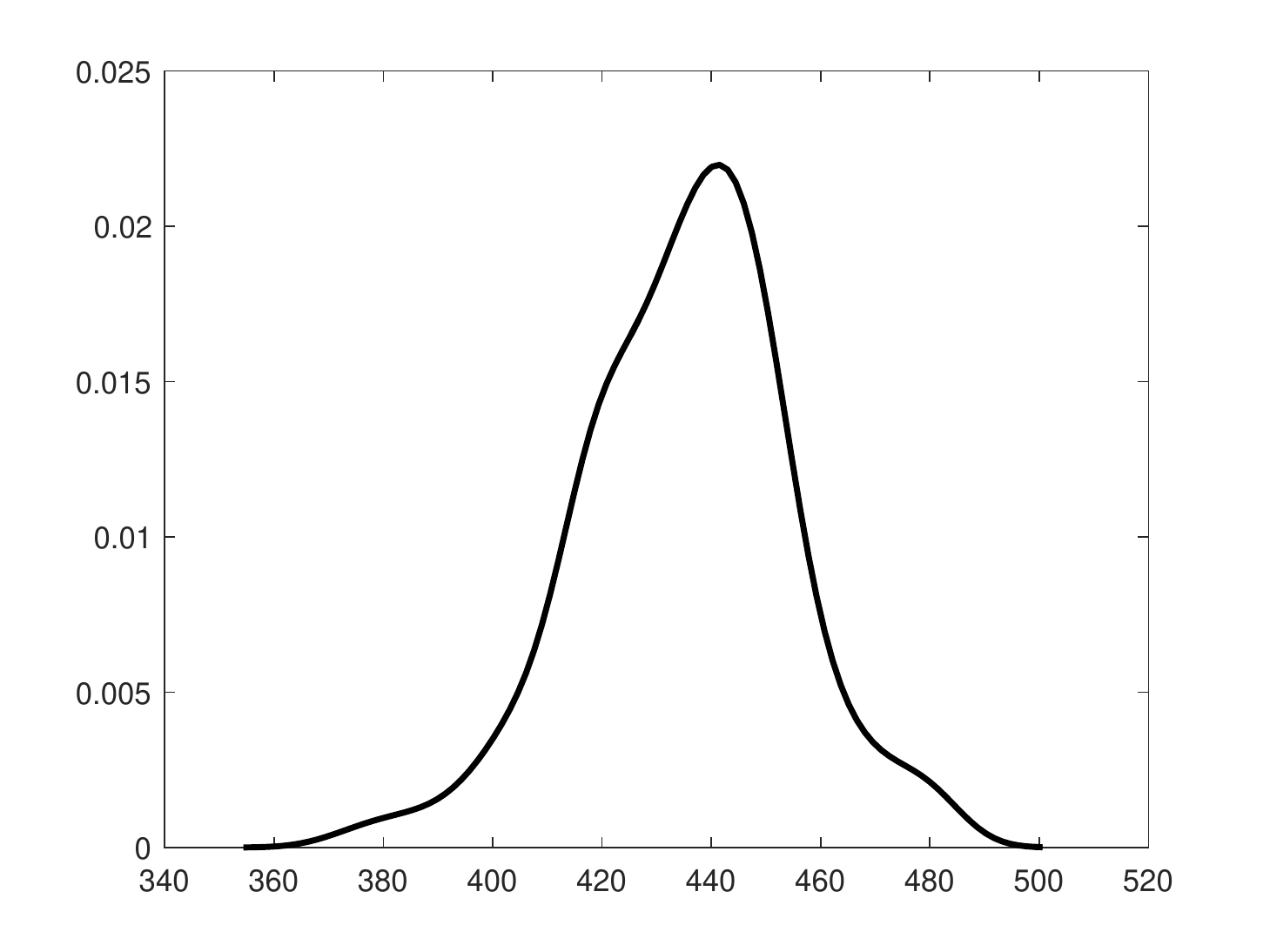}\label{figsub:ss14}}
	\subfigure[$c_6$]{\includegraphics[height=0.15\textheight,width=0.3\textwidth]{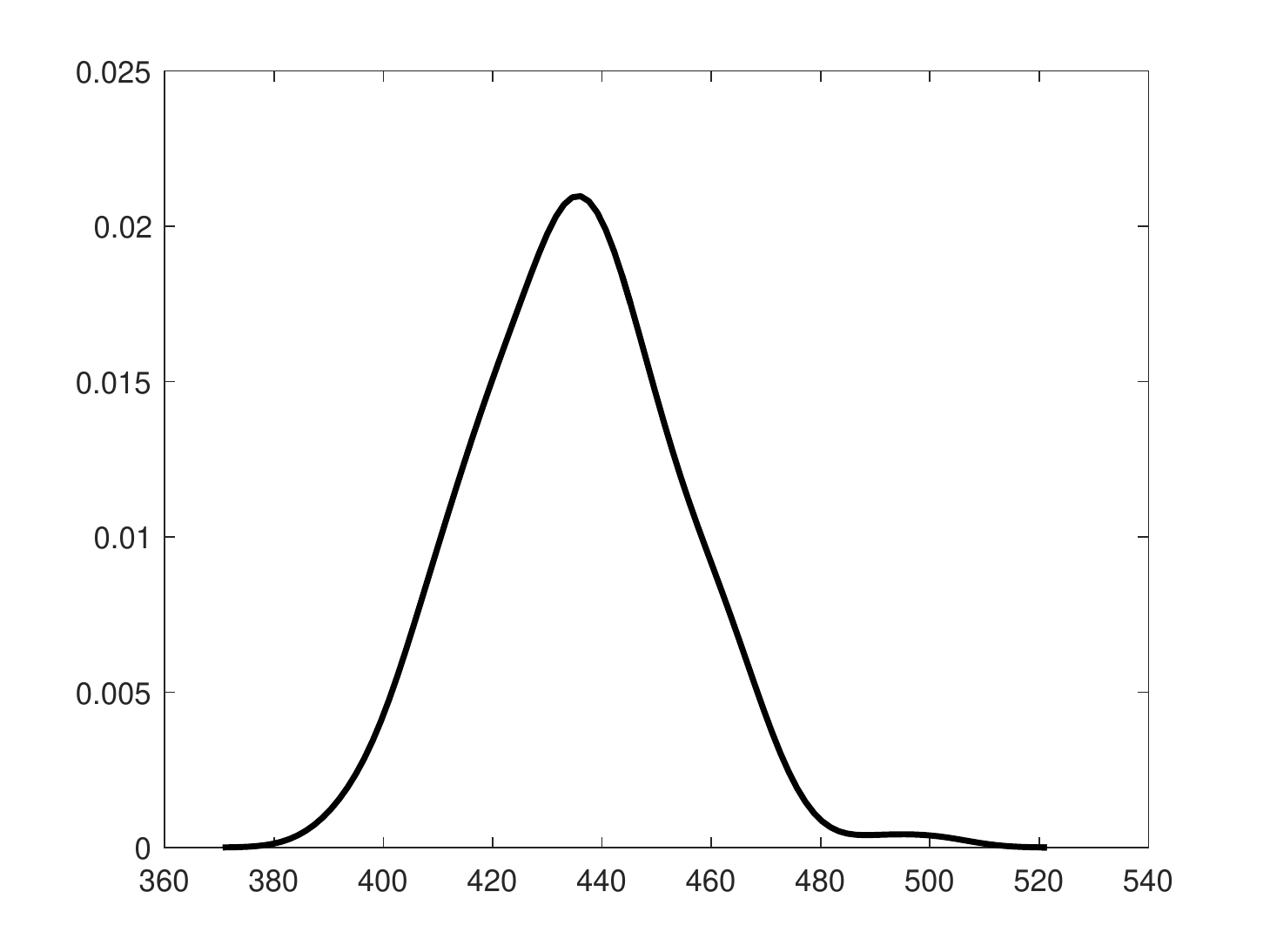}\label{figsub:ss15}}
	\caption{Estimated marginal distributions of the 15 summary statistics using $n=200$ for the melanoma cell biology applications when $P_m = 0.1$, $P_p = 0.0015$ and $q = 0.25$.}
	\label{fig:ss}
\end{figure}

The results, in comparison to those obtained in \citet{Vo2015}, are shown in Figure \ref{fig:melanoma_sim} for the simulated data and Figure \ref{fig:melanoma} for the real data.  From Figure \ref{fig:melanoma_sim} it can be seen that BSL approaches produced results similar to that of ABC for the simulated data.  It appears that BSL is able to accommodate the 15 summary statistics directly without further dimension reduction.  However, it is clear that the dimension reduction procedure of \citet{Vo2015} performs well.  From Figure \ref{fig:melanoma} (real data) it is evident that ABC and the BSL approaches produce similar posterior distributions for $P_p$  and $q$.  For $P_m$, there is a difference of roughly 0.01 between the posterior means of the BSL and ABC approaches and an increase in precision for BSL.  This discrepancy for the real data not apparent in the simulated data requires further investigation.  One potential source of error for BSL is the multivariate normal assumption.  The estimated marginal distributions of the summary statistics (using $n=200$) when the parameter is $\theta = (0.1, 0.0015, 0.25)$ is shown in Figure \ref{fig:ss}.  All distributions seem quite stable but there is an indication of non-normality for some of the summary statistics.  Given the results in Figure \ref{fig:melanoma_sim} and \ref{fig:melanoma}, it appears that BSL is showing at least some robustness to this lack of normality.  

\section{Further Reading} \label{sec:Further}

\citet{Ong2016} developed a stochastic optimisation algorithm to obtain a variational approximation of the BSL posterior.  The authors utilise an unbiased estimator of the log of the multivariate normal density due to \citet[pp.\ 56]{Ripley1996}.  \citet{Ong2016} demonstrated that significant computational savings can be achieved relative to MCMC BSL, at the expense of resorting to a parametric approximation of the posterior.  This work has been extended by \citet{Ong2017} to higher dimensional summary statistic and parameter spaces.

\citet{An2016} and \citet{Ong2017} considered shrinkage estimators of the covariance matrix of the model summary statistic in order to reduce the number of simulations required to estimate the synthetic likelihood.

\citet{Pham2014} replaced the ratio of intractable summary statistic likelihoods of the Metropolis-Hastings ratio in an MCMC algorithm with the outcome of a classification algorithm.  Datasets are simulated under the current parameter and proposed parameter with the former observations labelled as class 1 and the latter labelled as class 2.  A classifier, \citet{Pham2014} used random forests, is then applied.  From the fitted classifier, the odds for the value of the observed summary statistic $s_{obs}$ is computed and used as a replacement to the ratio of intractable likelihoods.  \citet{Pham2014} noted that BSL is a special case of this approach when classical quadratic discriminant analysis is adopted as the classifier.

\citet{Everitt2015} suggested that the SL can be used to perform Bayesian model selection in doubly intractable models, which contain an intractable normalising constant that is a function of $\theta$.  Such models can occur in complex exponential family models such as exponential random graph models for social networks and the Potts model for image analysis.  \citet{Everitt2015} developed computational algorithms in order to produce an SL approximation to the evidence $p(s_{obs}) = \int_{\theta}p(s_{obs}|\theta)\pi(\theta)d\theta$ for each model.

\section{Bayesian empirical likelihood}

ABC is a popular computational method of choice not only when there is no likelihood, but also when the likelihood is available 
but difficult or impossible to evaluate. 
Another popular idea is to replace the likelihood itself with an empirical alternative. 
This so-called empirical likelihood (EL) can be embedded within an ABC algorithm or provide an alternative to ABC.
The approach is appealing even for less complex models if there is a concern that the model is poorly specified. 
For instance, if the likelihood is a mixture but is misspecified as a single normal distribution, the corresponding parameter estimates, intervals and inferences
may exhibit unacceptably poor behaviour \citep{Chen2003}. In this case, normal approximation confidence intervals perform poorly in the area of interest, i.e. the lower tail, but intervals based on an EL 
are shown to perform as well as intervals based on a correctly specified mixture model.  

EL has been shown to have good small sample performance compared with methods that rely on asymptotic normality. Moreover, it enables
distribution-free tests without simulation, and provides confidence intervals and regions that have appealing theoretical and
computational properties \citep{Owen1988, Owen2001}. Some of these features are discussed in more detail below.

Close parallels with the EL approach have been drawn with estimating equations \citep{Qin1994, Grendar2007}, 
kernel smoothing in regression \citep{Chen2009a, Chen2009b, Haardle1990, Fan1996}, 
maximum entropy \citep{Rochet2012} and functional data analysis \citep{Lian2012}. We do not elaborate on these associations in 
this chapter, but refer the interested reader to the cited references.

EL approaches have been developed in both frequentist and Bayesian contexts. 
This section provides a brief overview of the method under both paradigms. We then focus on a particular algorithm, $BC_{el}$, 
proposed by \citet{Mengersen2013}, which was first conceived as part of an ABC algorithm but was then developed independently of the
ABC architecture.

\subsection{Empirical Likelihood}
\label{sub:emplik}

Empirical likelihood (EL) has been a topic of active research and investigation for over a quarter of a century. Although similar ideas were established earlier 
(see, for example, the proposal of a ``scale-load'' method for survey sampling by \citet{Hartley1968}), 
EL was formally introduced by \citet{Owen1988} as a form of robust likelihood ratio test. 

Following \citet{Owen1988} and \citet{Owen2001}, assume that we have i.i.d. data $Y_i, i=1,...,n$ from a distribution $F$. 
An EL denoted $L(F)$ is given by
$$ L(F) = \prod_{i=1}^{n} F( \{y_i\} )  . $$
The likelihood ratio and corresponding confidence region are given by, respectively,
$$ R(F) = L(F) / L(\hat{F}) \quad \mbox{and}$$  
$$ \{T(F) | R(F) \geq r \} $$
\noindent where $\hat{F}$ is the empirical distribution function and for some appropriate value of $r$.

Given parameters of interest $\theta$ and an appropriate sufficient statistic $T(F)$ for it, a profile likelihood and corresponding confidence region become, respectively,
$${\mathcal{R}}(\theta) = {\rm sup} \left\{ R(F) | T(F) = \theta \right\} \quad \mbox{and}$$
$$ \{ \theta | {\mathcal{R}}(F) \geq r \}. $$
If there are no ties, we let $p_i = F(\{y_i\})$, $p_i>0$, $\sum_{i=1}^{n}p_i=1$, and find that
$$L(F) = \prod_{i=1}^{n} p_i  \ \ ; \ \ L(\hat{F}) = \prod_{i=1}^{n} 1/n$$
$$R(F) = \prod_{i=1}^{n} np_i \ \ ; \ \ {\mathcal{R}}(\theta) = {\rm sup} \left\{ \prod_{i=1}^{n} np_i | T(F) = \theta \right\} \ .$$
Obvious adjustments are made to $L(F)$ and $L(\hat{F})$ if there are ties.

A fundamental result obtained by \citet{Owen1988} is that if the mean $\theta_0$ of the distribution $F$ is finite and its covariance matrix is finite with 
rank $q>0$, then as $n\rightarrow\infty$, 
$$-2 \log R(\theta_0) \rightarrow \chi_q^2 .$$
This is the same as that obtained by Wilks' Theorem for the parametric setup.
Thus for a $100(1-\alpha)$\% confidence region, 
$r = r_0 = \exp (-\chi^2_{q,\alpha/2})$.

As a concrete example of EL, suppose that interest is in estimation of the mean, i.e., $\theta= E[Y]$. Then
$T(\hat{F}) = n^{-1} \sum_{i=1}^{n} y_i$, with confidence region and profile likelihood given by, respectively,
$$ \left\{ \sum_{i=1}^{n} p_i y_i | p_i\geq 0, \sum_{i=1}^{n} p_i=1, \prod_{i=1}^{n} np_i>r\right\}  \quad \mbox{and}$$
$$ R(\theta) = {\rm sup} \left\{ \prod_{i=1}^{n} np_i | p_i > 0, \sum_{i=1}^{n} p_i=1, \prod_{i=1}^{n} np_i=\theta\right\} \ .$$
Thus under certain conditions, a $(1-\alpha)$-level EL confidence interval for $\theta_0=\bar{Y}$ is given by
$$ \{\theta | r(\theta) \leq \chi_1^2(\alpha))$$ 
where $r(\theta) = -2 \sum \log(n\hat{p}_i)$ is the log EL function and $\chi_1^2(\alpha)$ is the upper $\alpha$ 
quantile of the $\chi^2$ distribution with one degree of freedom. 

The above set-up can also be seen as an estimating equation problem, where the true value $\theta_0$ satisfies the estimating
equation 
$$E_F[m(Y;\theta)] = 0 \ $$
with $m(Y;\theta)$ denoting a vector-valued (estimating) function. Hence we can take $m(Y;\theta)=Y-\theta$ to indicate a vector mean,
$m(Y;\theta)=I_{Y\in A}-\theta$ for $Pr(Y\in A)$, $m(Y;\theta)=I_{Y<\theta}-\alpha$ for the $\alpha$th quantile of $Y$ if
$Y$ is continuous, $m(Y;\theta)=I_{Y\leq\theta}-0.5$ for the median, and so on.

More generally, we have one or more constraints of the form  $E_F[h(Y,\theta)]=0$,
where the dimension of $h$ sets the number of constraints in unequivocally defining the parameters of interest $\theta$. 
Then the EL is defined as
$$L_{el}(\theta|y) = \max_p \prod_{i=1}^{n} p_i$$
for $p \in [0,1]^n$, with constraints 
$$\sum_{i=1}^n p_i = 1;     \sum_{i=1}^n p_i h(y_i, \theta) = 0 \ . $$ 

Perhaps surprisingly, there are relatively few Bayesian formulations of EL in the published literature. 
An earlier Bayesian ABC approach using an approximation of the EL based on the pairwise score equation was 
proposed by \citet{Pauli2010}. 
The authors focused on establishing the validity of the procedure, arguing that its asymptotic properties were preferred over 
the approximations employed by \citet{Pauli2011}. See also \citet{Ruli2015}. 
\citet{Owen2001} (Ch.\ 9) noted some parallels between EL and the Bayesian bootstrap \citep{Rubin1981}, and 
\citet{Rochet2012} has suggested a Bayesian approach to generalised empirical likelihood, and generalised method of moments, via a 
form of maximum entropy. \citet{Chaudhuri2011} describe Bayesian EL approaches in a spatial modelling context, as discussed in more
detail below. 

More direct research into Bayesian EL comprise a Monte Carlo study \citep{Lazar2003} and two probabilistic 
studies \citep{Schennach2005, Grendar2007}. 
In contrast to the reported Bayesian bootstrap-type approaches of \citet{Owen2001}, 
\citet{Schennach2005} and \citet{Rasuga}, \citet{Grendar2007, Grendar2009} proposed a Bayesian large deviations (law of large numbers) 
probabilistic interpretation and justification of EL. They showed that, in a parametric estimating equations setting, the EL method 
is an asymptotic instance of the Bayesian non-parametric maximum a-posteriori approach. 

\subsection{Features of EL}
\label{sub:features}

Since Owen's paper in 1988, the properties of EL have been comprehensively investigated and reviewed \citep{Hall1990, Owen2001}. 

EL methods have been favourably contrasted with common alternatives for estimation of complex models. For example, a natural competitor is calibration, which 
proceeds by choosing, by some method, parameter values that match selected features of the observed data. This can be difficult for richly parametrised models 
with strong correlation structure. EL can be perceived as a more statistically formal method of calibration in that it uses moments for matching. 
Another common competitor, maximum likelihood, requires the definition, estimation and maximisation of a likelihood and 
can be both analytically and computationally demanding for complex models. In contrast, EL requires only summary (moment) statistics 
and can perform inference on an approximate likelihood, but inherits the properties of standard likelihood \citep{Owen2001}. 
These properties of standard likelihood are principally obtained by appeal to Wilks' Theorem \citep{Qin1994}. 

As described above, likelihood ratio confidence regions can be constructed by EL that often do not require 
estimation of the variance \citep{Chen2009a, Chen2009b} and have the same 
order of magnitude error as their parametric counterparts. This also applies for more general regression contexts 
\citep{Chen1993, Chen1994, Chen2006, Chen2007}. The confidence regions constructed in this manner respect the boundaries of 
the support of the target parameter and are more natural in shape and orientation of the data since they contour a
log-likelihood ratio. In particular, they are often superior to confidence regions based on asymptotic normality when 
the sample size is small. The confidence regions can be further improved by applying 
Bartlett's correction, $(1_a/n)\chi^2_{q,\alpha}$, where $a$ involves higher order moments of $Y$ \citep{DiCiccio1991}.

A key assumption underlying standard EL is that the random variables are independent with a common distribution. 
An analogue, the weighted EL or exponentially tilted distribution accommodates data that are independent 
but not necessarily identically distributed. This approach was introduced by \citet{Schennach2005} and has been taken up by a
large number of authors \citep{Owen2001, Kitamura2006, Glenn2007}.   
\citet{Chaudhuri2011} contrast the two approaches as follows. They frame the EL as
$$ l(\theta) = \prod_{i=1}^{m} \hat{w}_i(\theta) $$
where
$$\hat{w}(\theta) = {\rm arg} \max_{w\in \mathcal{W}_0} \sum_{i=1}^m f\{w_i(\theta)\}$$
for some specified function $f$. 
They then consider standard EL as a form of constrained maximum of a nonparametric 
likelihood since for a given $\theta$, $l(\theta)$ equals the EL when $f(w_i) = \log(w_i)$,
and the exponentially tilted likelihood as a form of maximum entropy such that $f(w_i)=-w_i\log (w_i)$.
As these authors discussed, the exponentially tilted likelihood can also be seen as a profile likelihood for $\theta$. 

Moreover, \citet{Schennach2005} shows that this reformulation of the maximisation problem of the EL allows for a probabilistic interpretation which justifies its use in a Bayesian setting. More precisely, the posterior distribution for a parameter of interest $\theta$ may be seen as

\begin{equation*}
\pi(\theta|y)=\pi(\theta) \int_\Psi\ L(\theta,\psi|y) \pi(\psi|\theta) d\psi ,
\end{equation*}
where $\psi$ represents a (potentially infinite-dimensional) nuisance parameter which absorbs all those aspects of the model not described by the parameter of interest $\theta$.  The information contained in the nuisance parameter may be discretised by a vector of parameter $\xi=(\xi_1,\ldots,\xi_N)$ with $N\rightarrow \infty$. The nuisance parmater $\xi$ may then be given a prior which shares the Dirichlet prior's property of providing posteriors which assign probability one to distributions supported by the sample. \citet{Schennach2005} shows then this reformulation has a computationally convenient representation, for which the posterior of the parameter of interest $\theta$ may be obtained through

\begin{equation*}
\pi(\theta|y)=\pi(\theta) \prod_{i=1}^n p_i^{\star}
\end{equation*}

\noindent where $p^\star=(p_1^{\star},\cdots,p_n^{\star})$ are the weights obtained as solution of the maximization problem

\begin{equation*}
L_{\mathrm{BETEL}}(\theta)=\max_{p^{\star}} \sum_{i=1}^{n} p_i^{\star} \log p_i^{\star}
\end{equation*}

\noindent under constraints $p^{\star} \in [0,1]^n, \sum_{i=1}^n p^{\star}_i = 1,     \sum_{i=1}^n p^{\star}_i h(y_i, \theta) = 0$, where ``BETEL'' stands for ``Bayesian exponentially tilted likelihood''. This method may be called ``Bayesian exponentially tilted EL'', because it uses the exponential tilting proposed in \citet{Efron1981} and has a Bayesian interpretation. This version of the EL will be used in Section \ref{sub:extension}. 

\citet{Glenn2007} examined the robustness properties of the estimates arising from the tilted distribution. 
For example, whereas the root mean squared error (RMSE) of the EL estimator for the mean increases as the non-iidness of the sample increases, 
the RMSE of the weighted EL estimator remains closer to its theoretical value.
Other extensions to standard EL, such as the continuous updating estimator, have also been proposed \citep{Hansen1996}.

In a Bayesian framework, the standard and exponentially tilted likelihoods have been shown to be appropriate for Bayesian inference 
for a range of set-ups and under certain conditions on the prior, particularly for a prior with sufficiently large variance 
\citep{Monahan1992, Lazar2003, Chaudhuri2011}.

Notwithstanding these attractions, there are some drawbacks in applying EL. One substantive issue is the formulation of the estimating equations. 
The number of equations is one issue: there should be at least as many as the dimension of the parameter space, but any more than this (which may be available and 
desirable from the perspective of model description) has been argued to adversely affect inference \citep{Qin1994}. However, it is suggested by
\citet{Mengersen2013} that this concern may not apply in all circumstances, in a Bayesian setup; this is illustrated in the $g$-and-$k$ example
given below.

\subsection{Estimation}

The most common approach to estimation of the EL is through the method of Lagrange multipliers.
In general terms, this method aims to maximise $f(x)$ subject to a (multivariate) constraint $g(x)=0$. This is
achieved by finding $x^*=x^*(\lambda)$ maximising $f(x)=\lambda^{\prime} g(x)$ such that $g(x^{\prime})=0$.
Then $x^*$ solves the constrained problem. 
Considering again the example of estimating $\theta=E[Y]$, the aim is to maximise 
$${\rm log}R(p_1,..,p_n) = \sum_{i=1}^{n} {\rm log}(np_i)$$
under the constraints
$$n\sum_{i=1}^{n} p_i(Y_i-\theta)=0,\quad  1-\sum_{i=1}^{n}p_i=0 \ .$$
We write 
$$G=\sum_{i=1}^{n}\log(np_i) - n\lambda\sum_{i=1}^{n}(Yi-\mu) - \gamma(1-\sum_{i=1}^{n}p_i)$$
where $\lambda$ and $\gamma$ are the Lagrange multipliers. This can be solved to find a unique solution for $\lambda=\lambda(\theta)$.

There is a range of software for computing the EL, particularly targeted towards specific applications. 
A helpful repository and description of available code is on Art Owen's website.
A powerful library available in the R software is the package `emplik' \citep{Zhou2014}. The underlying computational method is based on the 
Newton-Lagrange algorithm, whereby the Lagrangian function described above is solved by an application of Newton's 
method, which iterativly uses a second order Taylor approximation of $f(x)$ to find an optimal value $x^*$ satisfying $f^\prime(x^*)=0$.

For example, the package {\tt el.test} in the {\tt emplik} library conducts a simple EL ratio test that returns $-2$ log likelihood 
ratio ({\tt-2LLR}, which has an approximate chi-squared distribution under the null hypothesis), the associated p-value, the final
value of the Lagrange multiplier ({\tt lambda}), the gradient at the maximum ({\tt grad}), the hessian matrix ({\tt hess}), weights on the 
observations ({\tt wts}) and the number of iterations performed ({\tt nits}). 

The following code, provided in the emplik documentation, illustrates two tests on a two-dimensional set of data: 
(i) $H_0: \mu_1 = \mu_2$ and  (ii) $H_0: 2\mu_1 - \mu_2 = 0$.

\begin{verbatim}
# generate data
x <- matrix(c(rnorm(50,mean=1), rnorm(50,mean=2)), ncol=2,nrow=50) 
y <- 2*x[,1]-x[,2]
# test hypothesis (i)
el.test(x, mu=c(1,2))
# test hypothesis (ii)
el.test(y, mu=0) 
\end{verbatim}

In one realisation of this code, the results of the first test were returned after four iterations, 
with weights ranging between 0.75 and 1.51, and with -2LL=1.50 and a p-value of 0.47 
under the assumption that $-2$LL is approximately chi-square under $H_0$.
The second null hypothesis returned a p-value of 0.22.

Examples of the use of the {\tt emplik} library for survival analysis are given by \citet{Zhou2015}. Whereas {\tt el.test} requires
uncensored data, the packages developed by Zhou and embedded in the {\tt emplik} library enable estimation of 
hazard functions, cumulative distribution functions and confidence bands for various types of censored 
data under a range of survival models. 

As an example, the package {\tt em.cen.EM} can be used to test the hypothesis
$H_0:  \int g(t) dF(t) = \mu$ versus  $H_a: \int g(t) dF(t) \neq \mu$, 
where $g(t)$ is a user supplied function. For instance, $H_0$ can be the test about 
the Kaplan-Meier mean and $g(t) = t$. The myeloma code in the Appendix  illustrates this by testing $H_0:F(10)=0.2$ in
the {\rm myeloma} dataset incorporated in the {\tt emplik} library. The code also finds the upper and lower confidence
limit of a Wilks confidence interval.  
The output of this analysis provides a value -2LL and a corresponding p-value.




Bayesian EL methods are typically analysed by solving the EL using a Lagrange or similar method, then
generating observations from the posterior distribution of the parameters of interest by an MCMC
method. A more detailed description of this approach is given in the context of spatial modelling in the next section. An
alternative approach, $BC_{el}$, which employs the {\tt emplik} library to obtain the required likelihood values, is also detailed in a subsequent section.

\subsection{EL in practice}

The EL approach has been shown to be applicable in a broad range of contexts \citep{Qin1994}. For example, following its formulation for estimation in 
linear regression \citep{Owen1991}, it has been extended to nonlinear, generalised, parameric, nonparametric and 
semiparametric models with and without missing data and censoring, time series models and varying-coefficient models; see the review of \citet{Chen2009b} 
and the references therein. The approach has also been proposed for testing; see again \citet{Chen2009b}.  \citet{Einmahl2003} have proposed omnibus tests based on EL for 
a wide range of hypothesis tests, including symmetry, exponentiality, independence and change of direction. Tests for stochastic ordering using EL have been proposed
by \citet{Elbarmi2013} and \citet{Elbarmi2015}.

\citet{Chaudhuri2011} have proposed an EL approach for small area estimation and have suggested that the approach is also applicable to general random 
and mixed effects models. As the authors argue, EL overcomes the distributional assumptions of the more dominant parametric models as well as the linearity assumptions 
of the nonparametric models that have been proposed for this problem. In addition, EL avoids the need for resampling methods like jacknife and bootstrap to obtain 
mean squared error estimation. The authors' methodology is developed using a multivariate-$t$ prior for the parameter vector $\theta$ and both the regular and 
exponentially tilted formulations for the EL.

A Bayesian EL approach for constructing intervals for the analysis of survey data has been explored by \citet{Rao2010}. This work builds on the EL approaches for 
complex survey analysis proposed by \citet{Chen1999} and \citet{Wu2006}.
\citet{Rao2010} provides a clear exposition of EL methods for sample surveys. The authors set up the problem as one in which $N_t$ denotes the number of units $U=\{1,2,...,N_t\}$, in a finite population of size $N=\sum_{t=1}^T N_t$, that have the value $y_t^*$, and $n_t$
denotes the number of units in the sample having this value $y_t^*, t=1,...,T$. The sample data are then reduced to a set of so-called scale-loads $(n_1,n_2,...,n_T)^\prime, n_t\geq 0, n=\sum_{t=1}^{T}n_t$. Assuming a negligible sampling fraction, the likelihood
can be approximated by using the multinomial distribution with a log likelihood given by 
$$l(p) = \sum_{t=1}^{T} n_t {\rm log}(p_t)$$ 
with $p_t=N_t/N$, and the MLE of 
$$\bar{Y} = \sum_{t=1}^{T}p_ty^n$$ 
is the sample mean
$$\bar{y} = \sum_{t=1}^{T}\hat{p}_ty^{n^*}, \ \  \hat{p}_t=n_t/n .$$ 
The authors make the connection with the work of \citet{Chen1999} and argue that this `scale-load' approach is ``in the same spirit'' 
as EL as described by \citet{Owen1988}.

As described above, survival analysis is another area that lends itself naturally to EL. 
The popular Kaplan-Meier curve is a nonparametric estimator of the survival function $S(t)=P(T\geq t)$, where $T$ denotes the time to an event.
It is conceptually straightforward to see that $S$ can be estimated as a maximum EL estimator.
This field has been developed by a number of authors: see, for instance, \citet{Wang2001} for a general
exposition of the survival model, \citet{Murphy1997} for doubly censored data, \citet{Qin2001} for Cox modelling using EL, 
and \citet{Mckeague2002} for an EL approach to relative survival. The recent text by \citet{Zhou2015} provides an excellent
overview of the field as well as new models and computational algorithms, with associated R code to facilitate implementation.
A simple illustration of an EL approach to survival analysis is provided in the next section.

Recent years have also seen an increase in popularity of EL for spatial modelling.
\citet{Chaudhuri2011} pioneered a Bayesian EL approach for small area estimation. Their model can accommodate continuous
and discrete and area- and unit-level data, random and mixed effects, and the original and exponentially tilted empirical likelihoods.

A similar approach has also been proposed recently by \citet{Porter2015a} for this purpose. 
The so-called semiparametric hierarchical EL (SHEL) model can be applied to irregular lattices and 
irregularly spaced point-referenced data, and was shown to have improved mean squared prediction error compared 
with standard parametric analyses in a simulation study, a large community survey and a bird survey. In the SHEL model, EL is employed in an empirical data model, which 
is combined with a parametric process model that accounts for the spatial dependence through a rank-reduced intrinsic conditional autoregressive (ICAR) prior and, finally, with a model at the highest level of the hierarchy describing the parameter. 

A companion paper by the same authors \citep{Porter2015b} extends this work to a multivariate context,
with focus on the Fay-Herriot (FH) model which is a mainstay in small area estimation. The argument is made that this approach 
encompasses spatial correlation (via the FH model) but avoids the usual restrictive
Gaussian distributional assumptions (via EL). 

One of the fields in which EL has been prominent is economics and related fields. For example, \citet{Riscado2012}  promote the use of EL as a natural framework for 
estimation of dynamic stochastic general equilibrium (DSGE) models for macroeconomic analysis, since these models represent complex economic systems as a constrained 
optimisation problem and can be described as a set of moment conditions. The authors favourably compare EL with calibration and ML approaches, since the model parameters 
have complex correlation structures that hinder calibration and are typically characterised by nonlinear systems of difference equations that have no closed form and hence 
hinder ML. The likelihood is thus often approximated and then estimated (and maximised) using methods such as the Kalman filter and sequential Monte Carlo. 
The authors interpret the EL approach as mapping from the set of moment conditions to the stochastic processes of the economic variables, and then performing estimation 
by inverting that mapping. As discussed above, the importance and very often the difficulty of defining a set of ``good" moment conditions,
or constraints, is highlighted in this setting.

\subsection{The $BC_{el}$ algorithm}

A Bayesian EL algorithm was proposed by \citet{Mengersen2013}.
It was originally designed in the spirit of ABC, in that it avoids computation of the likelihood, but during its development the
authors realised that simulation from the likelihood could also be avoided and replaced with importance sampling.
Thus the so-called $BC_{el}$ algorithm generates values $\theta_i, i=1,\ldots,M$ from the prior distribution for $\theta$ and 
uses the values $w_i = L_{el}(\theta_i|y)$ as (unnormalised) weights in an importance sampling (IS) framework. The basic $BC_{el}$ sampler
is given below.  Of course, this IS algorithm will not be efficient if the posterior is very different to the prior.  Later, we describe a more sophisticated algorithm based on adaptive multiple importance sampling (AMIS, \citet{Cornuet2012}).

\begin{algorithm}
	\caption{$BC_{el}$, \citet{Mengersen2013}}
	\label{bcel}
	
	\begin{algorithmic}
		\FOR{$i = 1$ to $M$}
		\STATE Generate $\theta_{i}$ from the prior distribution $\pi(\cdot)$
		\STATE Set the (unnormalised) weight $\omega_{i}= L_{el}(\theta_{i}|y)$
		\ENDFOR
	\end{algorithmic}
	
\end{algorithm}


\subsubsection{Example 1}

As a concrete example, consider estimation of the population mean $\theta$ based on a sample of observations $y_i, i=1,...,n$.
In this case, two main decisions are required: the prior on $\theta$ and the constraints.
The computation of the EL $L_{el}(\theta_{i}|y)$ can be performed using the {\tt el.test} package in the 
{\tt emplik} library in {\tt R}, as described earlier in this chapter. In this case, the unnormalised weight $\omega_i$ is taken to be equal to 
the value of the empirical likelihood, which is calculated from the value of -2LLR obtained from the {\tt el.test} function. 

Suppose that a sample of size 100 is drawn from an (unknown) distribution $N(10,1)$ and the aim is to estimate the 
population mean $\theta$. A $N(-10,30)$ prior is imposed on $\theta$ and a first-moment constraint is chosen,
i.e., that the sample mean should equal the population mean. 
For the analysis, it is decided to run $M=5000$ iterations, noting that in practice a smaller value of $M$ can be used
but care must be taken to check that the weight has not concentrated too strongly on a small 
number of sampled values of $\theta$. A resampling step can be included to mitigate this, although at a cost of introducing additional variance. In this case, the algorithm becomes:

\begin{algorithm}
	\caption{$BC_{el}$ algorithm for Example 1}
	\label{example1}
	
	\begin{algorithmic}
		\FOR{$i=1$ to $M$}
		\STATE Generate $\theta_i \sim N(0,5)$
		\STATE Obtain $-2LL$ from {\tt el.test(y,mu=0)}
		\STATE Let $\omega_i =\exp(-0.5\times(-2LL))$ 
		\ENDFOR
		\STATE Resample $\theta$ with probability $\omega$ 
		\STATE Calculate summary statistics of the resampled values of $\theta$ \\
	\end{algorithmic}
	
\end{algorithm}


\noindent Example R code for this algorithm is given below.  

\begin{verbatim}
data = rnorm(100,10,1)
M = 5000; theta.propose=w=rep(0,M)
for (i in 1:M){
theta.propose[i] = runif(1,-10,30)
el = el.test(data,mu=theta.propose[i])
w[i] = exp(-0.5*(el$'-2LLR'))
}   
theta=sample(theta.propose,M,prob=w,replace=TRUE)
mean(theta); sd(theta); quantile(theta,probs=c(0.1,0.9))
hist(theta, main="",xlab="theta")
\end{verbatim}

\begin{figure}
	\centering
	\includegraphics[height=0.35\textheight,width=0.7\textwidth]{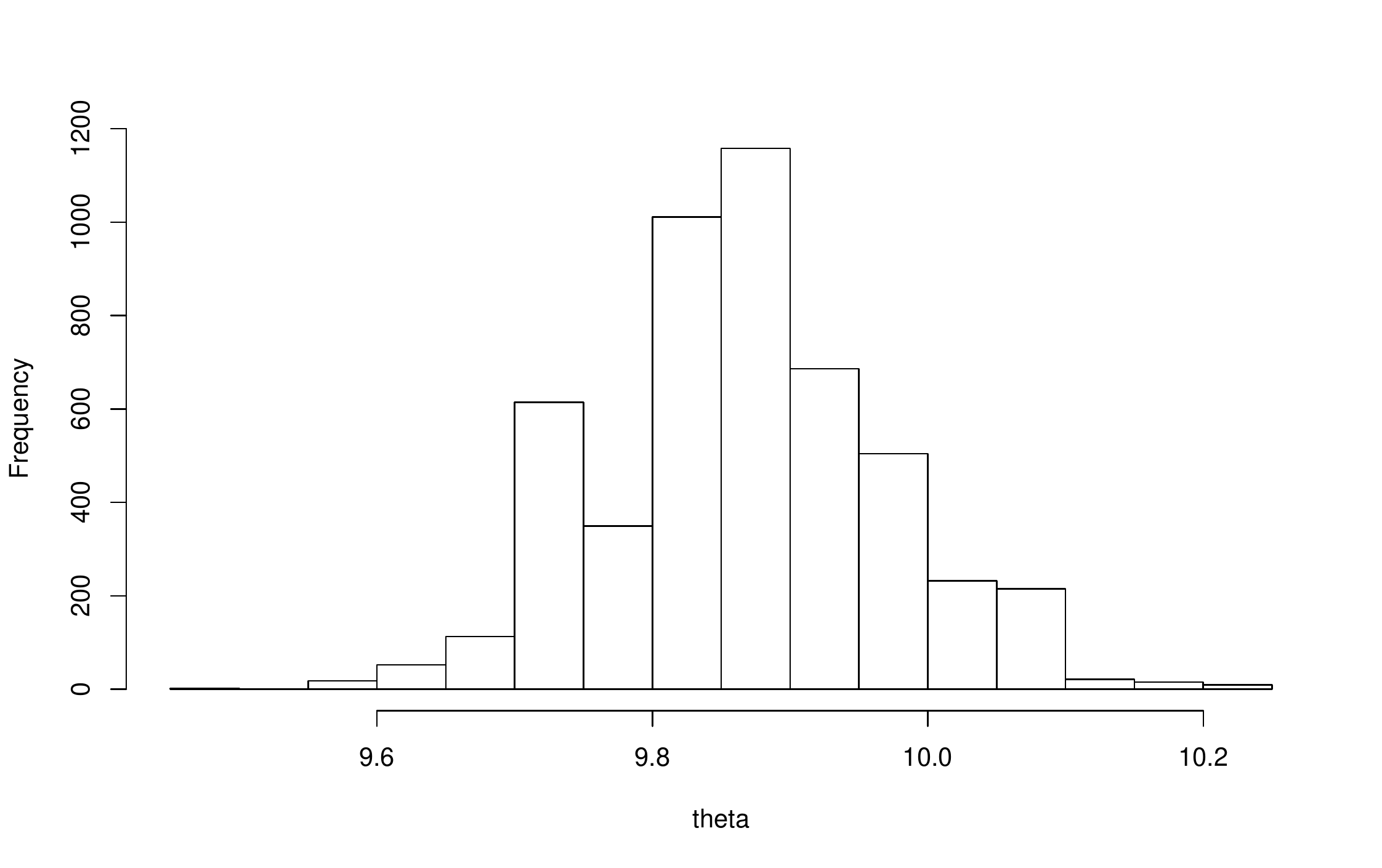}
	\caption{Histogram of draws from the $BC_{el}$ posterior distribution of $\theta$ based on data generated from a N$(\theta=10,1)$ distribution}
	\label{fig:n10hist}
\end{figure}

As noted above, the resampling step could be replaced with a weighted mean, standard deviation and quantiles.  One realisation of this code provided the following estimates: $\hat{\theta}=10.08$; s.d.$(\theta)=0.12$; 95\% CrI=$(9.88, 10.21)$.
A histogram of the obtained sample of $\theta$ is given in Figure \ref{fig:n10hist}.

\citet{Mengersen2013} comment on the performance of this algorithm with different constraints, namely one, two and three central
moments, $E[(X-\theta)]=0$, $E[(X-\theta)^2]=0$, $E[(X-\theta)^3]=0$. 
They observe that one and two constraints work well, but three constraints performed more poorly. This was seen to support the
general suggestion by \citet{Owen2001} that the number of constraints should be equal to the number of parameters.
Interestingly, this was not seen to be the case for the $g$-and-$k$ distributions, as described in the next example.

A possible measure of the efficiency of the algorithm is the effective sample size (ESS). 
The ESS is reportedly a measure of the `equivalent number of 
independent observations' in a sample, that is, the value that equates the obtained variance of the estimator of interest with the 
equivalent variance assuming an independent sample.
For weighted samples as in EL, the ESS can be estimated as 
$${\rm ESS} = 1/\sum_{i=1}^M \{ w_i / \sum_{j=1}^M w_j \}^{2} \ .$$  

\citet{Kish1965} argues that this substitution (of the EL for the exact likelihood) can 
be employed in any algorithm that samples from a posterior distribution. 
For example, it can be employed in composite likelihoods which are commonly used in areas such as population genetics
where the likelihoods are known but complex, and hence computationally difficult.  
The `traditional' composite likelihood approach decomposes the target distribution $\pi(\theta)L(\theta|y)$ into several 
multivariate Student $t$ distributions. In the $BC_{el}$ approach, the EL is used instead. 
The computation is achieved using AMSI, which can be parallelised on a multi-core computer.
The method can also be tailored for some non-i.i.d. problems such as dynamic models with AR structure, 
although the challenge here is in selecting appropriate constraints; 
see \citet{Mengersen2013} for details.

\subsection{Example 2}
\label{sub:quantile}

We illustrate the use of $BC_{el}$ by expanding on the discussion by \citet{Mengersen2013} of quantile distributions. 
These distributions are appealing for ABC in general, and $BC_{el}$ in particular: there is typically no closed form expression for the likelihood, 
so regular algorithms such as MCMC are not immediately applicable; and it is fast and easy to obtain simulations from a quantile function via an 
inversion algorithm.

There is a body of literature on using ABC for estimation of quantile distributions. \citet{Allingham} proposed an ABC-MCMC algorithm in which draws of the parameters of the quantile distribution are based on a Metropolis algorithm with a
Gaussian proposal distribution, and are accepted based on the rule $||D-D^\prime||<h$, where $D$
is the entire set of order statistics, $||\cdot||$ is the Euclidean norm and $h$ is heuristically chosen
after inspection of a histogram of  $||D-D^\prime||$ obtained from a preliminary run using a very large value
of $h$.  \citet{Peters2006} also developed an ABC-MCMC algorithm for complex quantile functions.  A range of improvements in the MCMC algorithm, 
selecting low-dimensional summary statistics and methods of choosing $h$ have since been suggested \citep{Prangle2011, Mcvinish2011}. 
Sequential Monte Carlo approaches for multivariate extensions of quantile distributions have also been proposed \citep{Drovandi2011}. 

The $g$-and-$k$ distribution is a popular example of a quantile distribution. This is a transformation of the standard normal distribution function, as follows:
$$ Q(z(p);\theta) = a + b\left(1 + c\frac{1-\exp(-gz(p))}{1 + \exp(-gz(p))}\right)(1 + z(p)^2)^kz(p) $$
where $\theta = (a,b,g,k)$ and $c$ is commonly set fixed at 0.8; see \citet{Rayner2002}.  
Here $p$ denotes the $p$th quantile from the $g$-and-$k$ distribution and $z(p)$ is the corresponding quantile of the standard normal distribution.  Thus simulation from the $g$-and-$k$ distribution requires only the generation of uniform($0,1$) variates. 

\begin{figure}
	\centering
	\includegraphics[height=0.35\textheight,width=0.5\textwidth]{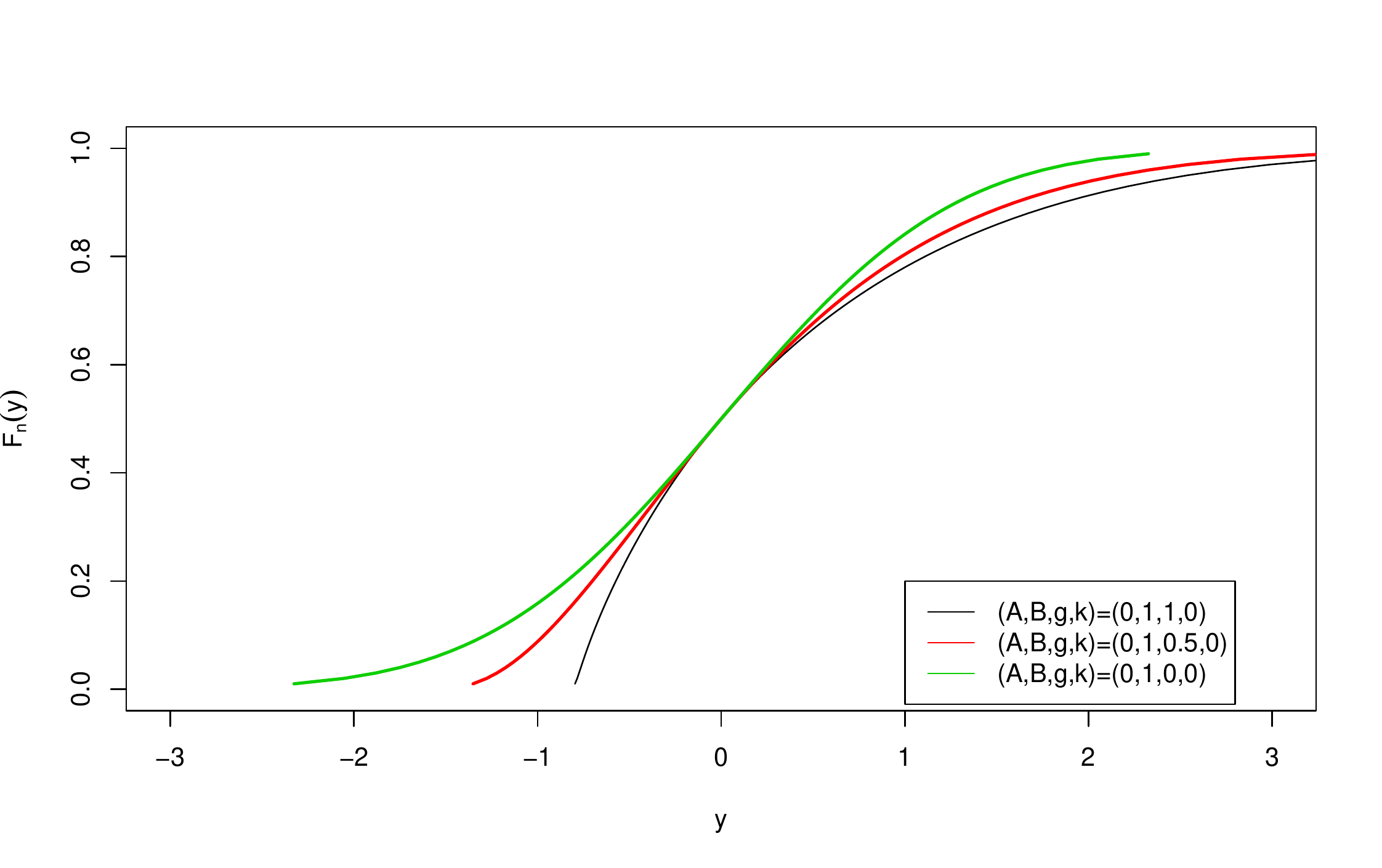}
	\caption{Cumulative distribution functions for three $g$-and-$k$ distributions.}
	\label{fig:gkcdf}
\end{figure}

Figure \ref{fig:gkcdf} shows the estimated cdf of a standard normal distribution based on a $g$-and-$k$ approximation, using the basic $BC_{el}$ procedure described in Algorithm \ref{bcel}. The parameters of the $g$-and-$k$ distribution corresponding to a $N(0,1)$ distribution are $\theta=(0,1,0,0)$. The analysis was based on $1000$ observations, 100,000 iterations and 10,000 resampled parameter values. The percentiles $(0.1, 0.25, 0.5, 0.75, 0.9)$ were chosen arbitrarily to form the constraints for EL and all parameters were generated from a $U[0,5]$ prior distribution.  

\begin{figure}
	\centering
	\includegraphics[height=0.35\textheight,width=0.7\textwidth]{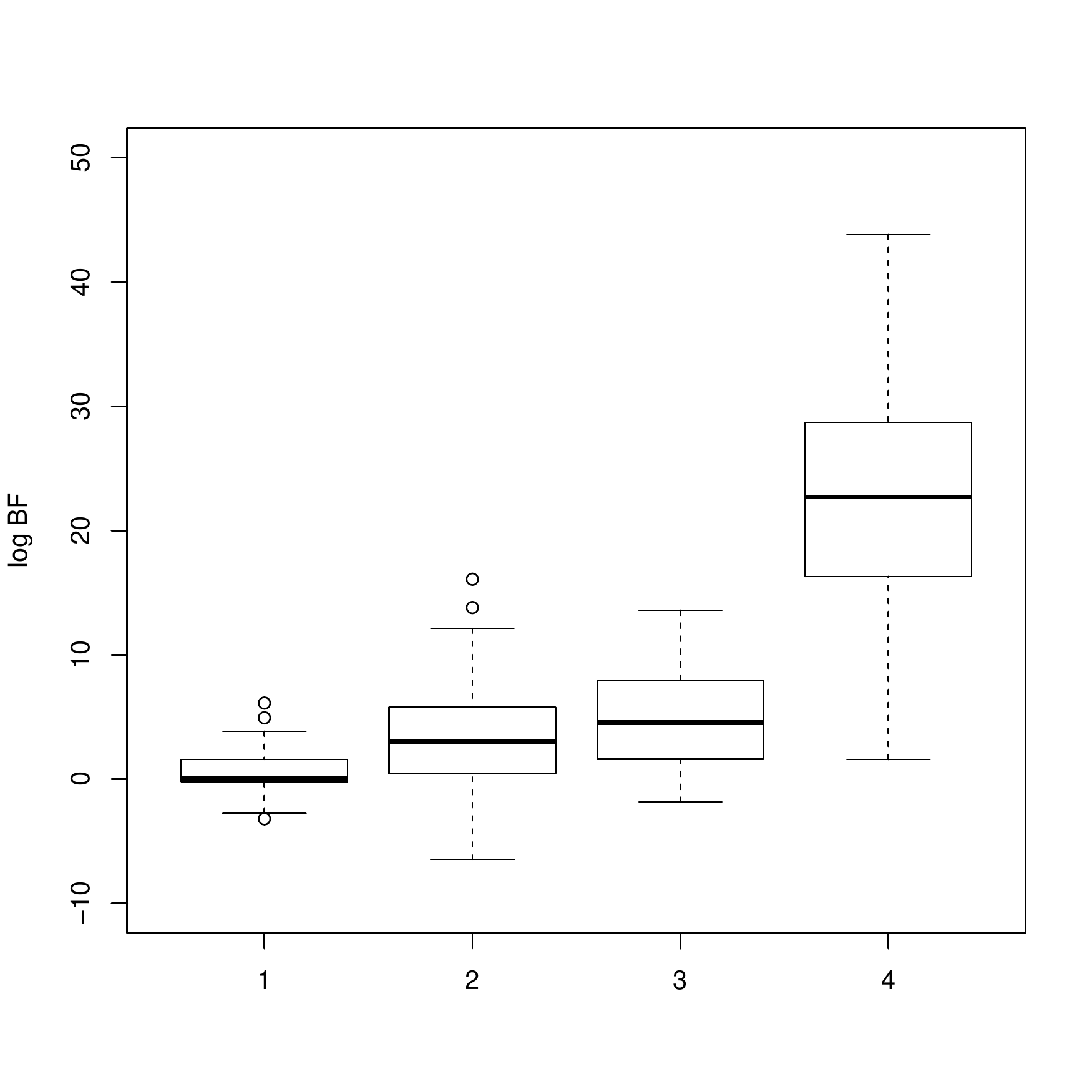}
	\caption{Boxplots of Bayes Factors comparing three $g$-and-$k$ distributions; data were generated from Model 1.  On the x-axis:  (1) Model 1 vs Model 2 with sample size $n=100$, (2) Model 1 vs Model 2 with sample size $n=500$, (3) Model 1 vs Model 3 with sample size $n=100$, (4) Model 1 vs Model 3 with sample size $n=500$.}
	\label{fig:gkBF}
\end{figure}

The Bayes Factor R code available in the Appendix illustrates the ease with which Bayes Factors (BF) can be computed for $g$-and-$k$ distributions using EL. The example
assumes a true model (Model 1) with $(A,B,g,k)=(0,1,1,0)$ versus two alternatives, $(0,1,0.5,0)$ (Model 2) and $(0,1,0,0)$ (Model 3).
Here, all models have zero mean ($A=0$) and unit variance ($B=1$) but differ in the degree of skewness, with Model 3 having
no skewness ($g=0$) and hence representing a standard normal distribution. 
The cumulative distributions functions for these three models are depicted in Figure \ref{fig:gkcdf}.
Two sample sizes of $100$ and $500$ and five constraints $(0.1, 0.25, 0.5, 0.75, 0.9)$ are considered. 
The resultant boxplots shown in Figure \ref{fig:gkBF} confirm that Model 1 is preferred over both of the alternative models, 
with a stronger log BF obtained for the larger sample size as anticipated.    

\subsubsection{Example 3}

\citet{Mengersen2013} also describe a variation on the basic $BC_{el}$ algorithm which employs AMIS 
in order to improve computational efficiency over plain importance sampling. The so-called $BC_{el}$-AMIS sampler employs multivariate 
Student $t_3(\cdot|m,\Sigma)$ distributions (3 degrees of freedom, mean $m$, covariance matrix $\Sigma$) as importance sampling 
distributions, as described in the following algorithm. The output of this algorithm is a weighted sample $\theta_{t,i}$ of size $MT_M$.  

\begin{algorithm}
	\caption{$BC_{el}$-AMIS}
	\label{algo:bcel-amis}
	
	\begin{algorithmic}
		\FOR{$i=1$ to $M$}
		\STATE Generate $\theta_{1,i}$ from the prior distribution 
		$\pi(\cdot)$. 
		\STATE Set the weight $\omega_{1,i }=L_{el}(\theta_{1,i}|y)$. 
		\ENDFOR
		\FOR{$t=2$ to $T_M$}
		\STATE Compute the weighted mean $m_t$ and weighted variance
		matrix  $\Sigma_t$ of the 
		$\theta_{s,i} (1\leq s\leq t-1, 1\leq i \leq M)$.
		\STATE Denote by $q_t(\cdot)$ the density of 
		$t_3(\cdot|m_t,\Sigma_t)$. 
		\FOR{$(i=1$ to $M)$}
		\STATE Generate $\theta_{t,i}$ from $q_t(\cdot)$. 
		\STATE Set $\omega_{t,i} = \pi(\theta_{t,i}) L_{el}(\theta_{t,i}|y) /
		\Sigma_{s=1}^{t-1} q_s(\theta_{t,i})$.  
		\ENDFOR
		\FOR{$(r=1$ to $t-1)$}
		\FOR{$(i=1$ to $M)$}
		\STATE Update the weight of $\theta_{r,i}$ as $\omega_{r,i} =
		\pi(\theta_{r,i}) L_{el}(\theta_{r,i}|y) /
		\Sigma_{s=1}^{t-1} q_s(\theta_{r,i})$. 
		\ENDFOR
		\ENDFOR
		\ENDFOR
	\end{algorithmic}
	
\end{algorithm}


\subsection{Extensions of the $BC_{el}$ algorithm}
\label{sub:extension}

Since its introduction, the $BC_{el}$ approach has been applied to a range of problems. For example, \citet{Cheng2014} cite the 
approach as the foundation for their proposed method for estimating the parameters of the extreme value model of \citet{Heffernan2004}.
Through a large simulation study, the method was found to provide good coverage of credible intervals, although one of the parameters needed more informative priors under some more challenging setups. 

In a second example, \citet{Grazian2016} discuss the use of $BC_{el}$ for copula estimation, whereby the marginal likelihood of the 
quantity of interest is approximated by the EL. 

Copula models are an important tool in multivariate analysis: while a huge literature exist about methods of estimating univariate marginal distributions, the problem of estimating the dependence structure of a multivariate distribution is more complex. Copula models allow for separately working with the univariate marginals and the joint distribution.  They are widely used in many applications, including actuarial sciences \citep{Frey2001}, epidemiology \citep{clayton78}, finance \citep{cherubini04}, hydrology \citep{salvadori07} among others. 

A copula model is a way of representing the joint distribution of a random vector $X= (X_1, \dots, X_d)$. Given a $d$-variate cumulative distribution function $F$ which depends on some parameter $\psi$, it is possible to show \citep{sklar} that there always exists a $d$-variate function $C_{\psi} : [0, 1]^d \to [0, 1]$, such that

\begin{equation*}
F(x_1,\dots , x_d;\lambda_1,\dots,\lambda_d,\psi) = C_{\psi}(F_1 (x_1;\lambda_1),\dots,  F_d (x_d;\lambda_d)), 
\end{equation*}

\noindent where $F_j$ is the marginal distribution of $X_j$, indexed by a parameter $\lambda_j$, and $\psi$ is a parameter characterizing the joint distribution.

In other terms, the copula $C$ is a distribution function with uniform margins on $[0, 1]$ which takes value from the univariate $F_1, F_2, \dots , F_d$ (which may be of the same form or may differ in terms of the parameters or of the forms) in order to produce the $d$-variate distribution $F$. The resulting model is very flexible, because it may utilise different types of marginal distributions and dependence structures. 

\begin{figure}
	\centering
	\includegraphics[height=0.3\textheight,width=1.0\textwidth]{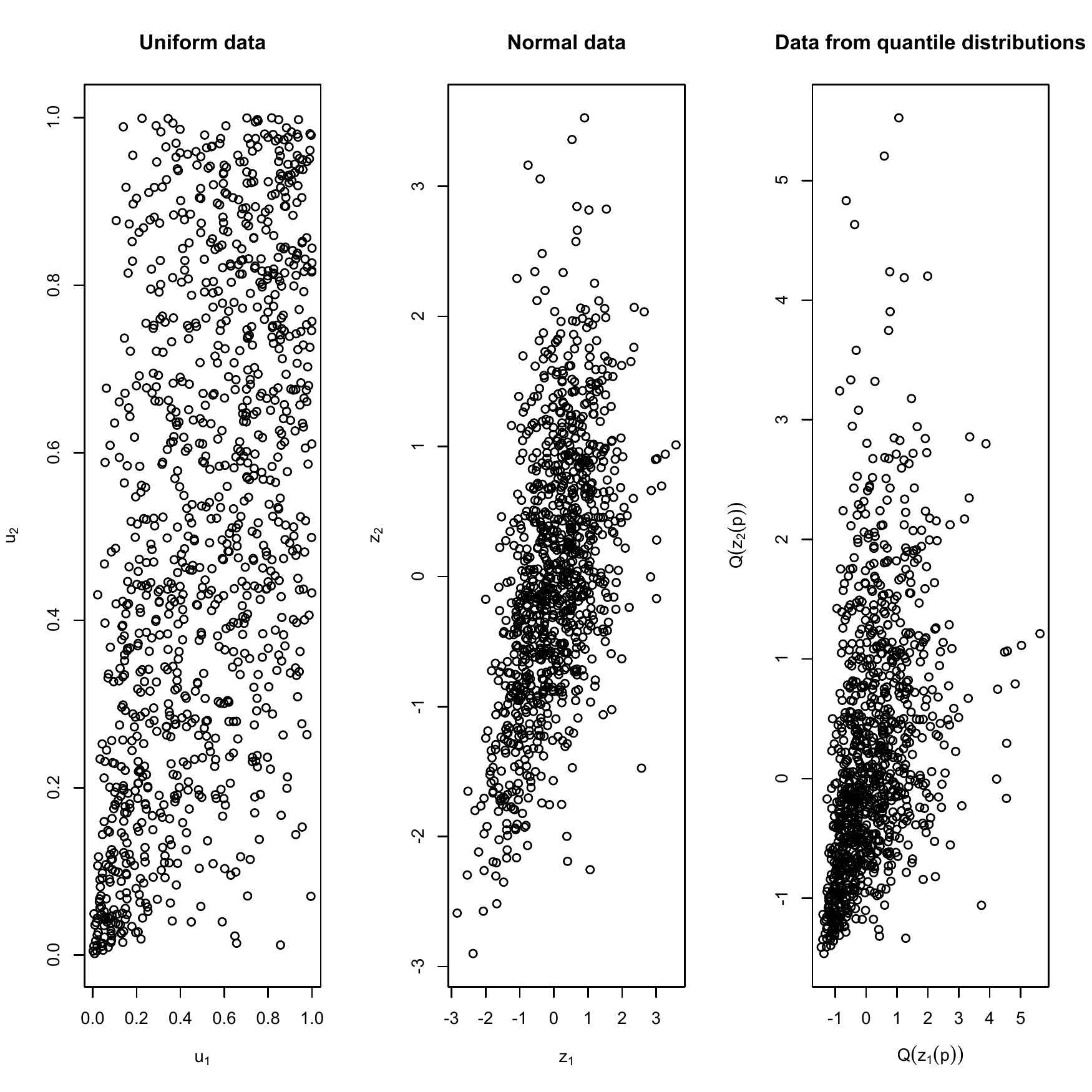}
	\caption{Scatterplots of the first two variables in the generation procedure: data from a Clayton copula with $\rho=0.5$ (left), transformed to normal data (centre) and, then, to data from a $g$-and-$k$ distribution with $a=0$, $b=1$, $g=0.5$ and $k=0$ (right).}
	\label{fig:copula_gen}
\end{figure}

Many different types of copula functions have been proposed in the literature; see \citet{Joe2015} for a review. An example is the Clayton copula, defined in the general $d$-dimension case as

\begin{equation*}
C(\mathbf{u}) = (u_1^{-\psi} + u_2^{-\psi} + \cdots + u_d^{-\psi} - d + 1)^{-\frac{1}{\psi}}
\end{equation*}

\noindent where $\psi \in [-1,\infty)\setminus \{0\}$ is a one-dimensional parameter. The Clayton copula is characterized by lower-tail dependence (that approaches $1$ as $\psi \rightarrow \infty$) and no upper-tail dependence. A representation of the Clayton copula (obtained through simulation) is available in Figure \ref{fig:copula_gen}. 

The frequentist standard method of estimating copula models is the ``inference from the margins'' (IFM) approach \citep{Joe2015}, i.e. a two-step procedure, where first the marginal distribution functions are separately estimated, either in a parametric or in a nonparametric way (depending on the information available on the marginals) and then the copula function is estimated. Bayesian alternatives have been explored, nevertheless they are still limited. The reader may refer to \citet{smith11} for a review. 

In some cases the interest of the analysis is in a function of interest $\theta$ of the copula and not in the complete dependence structure; this may be due to a weak information about the type of structure or to the need of a low-dimensional quantification of the dependence. Some typical quantities of interest are, for example, tail dependence indices, Spearman's $\rho$ or Kendall's $\tau$. While tail dependence indices represent, in the bivariate case, the probability that a random variable exceeds a certain threshold given that another random variable has already exceeded that threshold  \citep{grossmass2007}, Spearman's $\rho$ and Kendall's $\tau$ are measures of rank correlation, which are both expressible in terms of the copula $C$. For example, the Spearman's $\rho$ in the bivariate case is defined as

\begin{equation}
\label{eq:rho}
\rho = 12 \int_{[0,1]^2} C(u,v) \, du \, dv - 3 = 
12 \int_{[0,1]^2} uv \, dC(u,v)  - 3.
\end{equation} 

In this case, \citet{Grazian2016}, in the same spirit of the IFM method, propose to first estimate the marginal distributions and then study the interest measure of multivariate dependence with an approximate Bayesian approach based on an estimation of the likelihood of $\theta$ via EL (the authors use its Bayesian modification described in \citet{Schennach2005} and in Section \ref{sub:features}). In this way, it is possible to avoid the complete definition of the dependence structure (usually difficult to be determined) and elicite the prior distribution only for the quantity of interest, in order to reduce the bias derived from wrong distributional assumptions. Moreover their Bayesian approach avoids the loss of information of the IFM method and may be proved to be consistent. 

The $BCOP$ (``Bayesian computation for copulas'') algorithm follows and its final output will then be a posterior sample drawn from an approximation of the posterior distribution of the quantity of interest $\theta$ (see Algorithm \ref{BCOP}).

\begin{algorithm}[!h]
	\caption{$BCOP$ algorithm, \citet{Grazian2016}}
	\label{BCOP}
	Given a $n\times d$ data set $x=\{x_1,\ldots,x_n\}^\prime$ and marginal posterior samples $\{\lambda_1^{(s)},\ldots,\lambda_d^{(s)}\}$ for $s=1,\cdots,S$
	\begin{algorithmic}
		
		\FOR{$s=1, \ldots , S$}
		\STATE  Use the $s$-th row of the posterior simulation 
		$\{\lambda_1^{(s)}, \lambda_2^{(s)}, \dots , \lambda_d^{(s)}\}$ to create a matrix of uniformly distributed data $u^{(s)}_{ij}=F_j(x_{ij};\lambda^{(s)}_j)$
		$$
		u^{(s)}= \begin{pmatrix}
		u_{11}^{(s)} & u_{12}^{(s)} & \dots & u_{1d}^{(s)} \\
		u_{21}^{(s)} & u_{22}^{(s)} & \dots & u_{2d}^{(s)} \\
		\dots  & \dots  & u_{ij}^{(s)} & \dots  \\
		u_{n1}^{(s)} & u_{n2}^{(s)} & \dots & u_{nd}^{(s)} \\
		\end{pmatrix} .
		$$
		
		\ENDFOR 
		
		\STATE Given a prior distribution $\pi(\theta)$ for the quantity of interest $\phi$,
		
		\FOR{$b=1,\ldots,B$}
		\STATE Draw $\theta^{(b)}\sim \pi(\theta)$
		\FOR{$s=1,\ldots,S$}
		\STATE Compute $L_{BEL}(\theta^{(b)};u^{(s)})=\omega_{bs}$	
		\STATE Take the average weight $\omega_b=S^{-1}\sum_{s=1}^S \omega_{bs}$
		\ENDFOR
		\ENDFOR
		\STATE Output $(\theta^{(b)},\omega_{b}), b=1,\ldots,B$
		
	\end{algorithmic}
	
\end{algorithm}

This approach presents several advantages with respect to classical approaches to copula estimation. First, it may be applied to a generic dimension $d$, while in the literature there is a huge difference in terms of consistency results on the proposed estimators between the bivariate and the multivariate case. The authors have applied the $BCOP$ algorithm to a maximum dimension equal to $50$ with no loss of precision and with a reasonable computational expense (it has to be noted that the algorithm may be easily parallelised in the first step of estimation of the marginals). Second, the method provides a quantification of the error of estimation, not easily available in the classical approach (see \citet{schmid:schmidt:07} for the Spearman's $\rho$ and \citet{schmidt:stadtmuller:06} for the tail dependence indices). Third, it avoids the specification of the particular copula function which describes the dependence structure; this is particularly important in absence of information on it, since methods to select the copula function are not yet fully developed. 

Since the interest is in small dimension parameter (often only one measure of dependence), the choice of the constraints should be easy; unfortunately,  in practical applications these conditions might hold only asymptotically. This is the case, for example, of the Spearman's $\rho$: its sample counterpart $\rho_n$ is only an asymptotically unbiased estimator of $\rho$ so the moment 
condition is strictly valid only for large samples. 

\citet{Grazian2016} also apply the method to a real data-set based on the study of the dependence among five Italian financial institutes, where the returns are supposed to marginally follow a $GARCH(1,1)$ model with Student's $t$ innovations. They show how it is possible to obtain an approximated posterior distribution of the Spearman's $\rho$ of the financial asset returns of these institutes with Algorithm \ref{BCOP}.

As an application, consider the setting of Section \ref{sub:quantile}, where five sets of observations are simulated from $g$-and-$k$ identical but not independent quantile distributions with $a=0$, $b=1$, $g=0.5$ and $k=0$. The dependence structure is described by a multivariate Clayton copula \citep{McNeil2009} with true unknown multivariate $\rho$ equal to $0.5$. There are many ways to extend the bivariate Spearman's $\rho$ defined in \eqref{eq:rho} to the multivariate case and they are not in general equivalent; nevertheless it is often of interest in many fields of application to describe the dependence with a low-dimensional quantity, for example in the multivariate analysis of financial asset returns where there is the need to express the amount of dependence in a portfolio by a single number. Here, the following is considered

\begin{equation}
\label{eq:multi_rho}
\rho =\frac{
	\int_{[0,1]^d} \left( C({u}) - \Pi({u}) \right )d {u}} {\int_{[0,1]^d} \left( M({u}) - \Pi({u}) \right )d {u}} 
= h(d) \left\{ 2^d \int_{[0,1]^d}C({u})d {u}-1 \right\},
\end{equation}

\noindent where $M({u})=\min(u_1, u_2, \dots, u_d)$ is the upper Fr\'echet- Hoeffding bound, and $h(d) = (d+1)/ \{2^d-(d+1)\}$. For a review of the definitions of the Spearman's $\rho$ in the literature one may refer to \citet{schmid:schmidt:07}.

Uniform data have been generated from a multivariate Clayton copula with $\rho=0.5$ and then inverted in order to obtain data from the corresponding quantile distributions. Figure \ref{fig:copula_gen} shows the correlation between the first two sets of observations generated with this procedure. 

\begin{figure}
	\centering
	\includegraphics[height=0.35\textheight,width=0.5\textwidth]{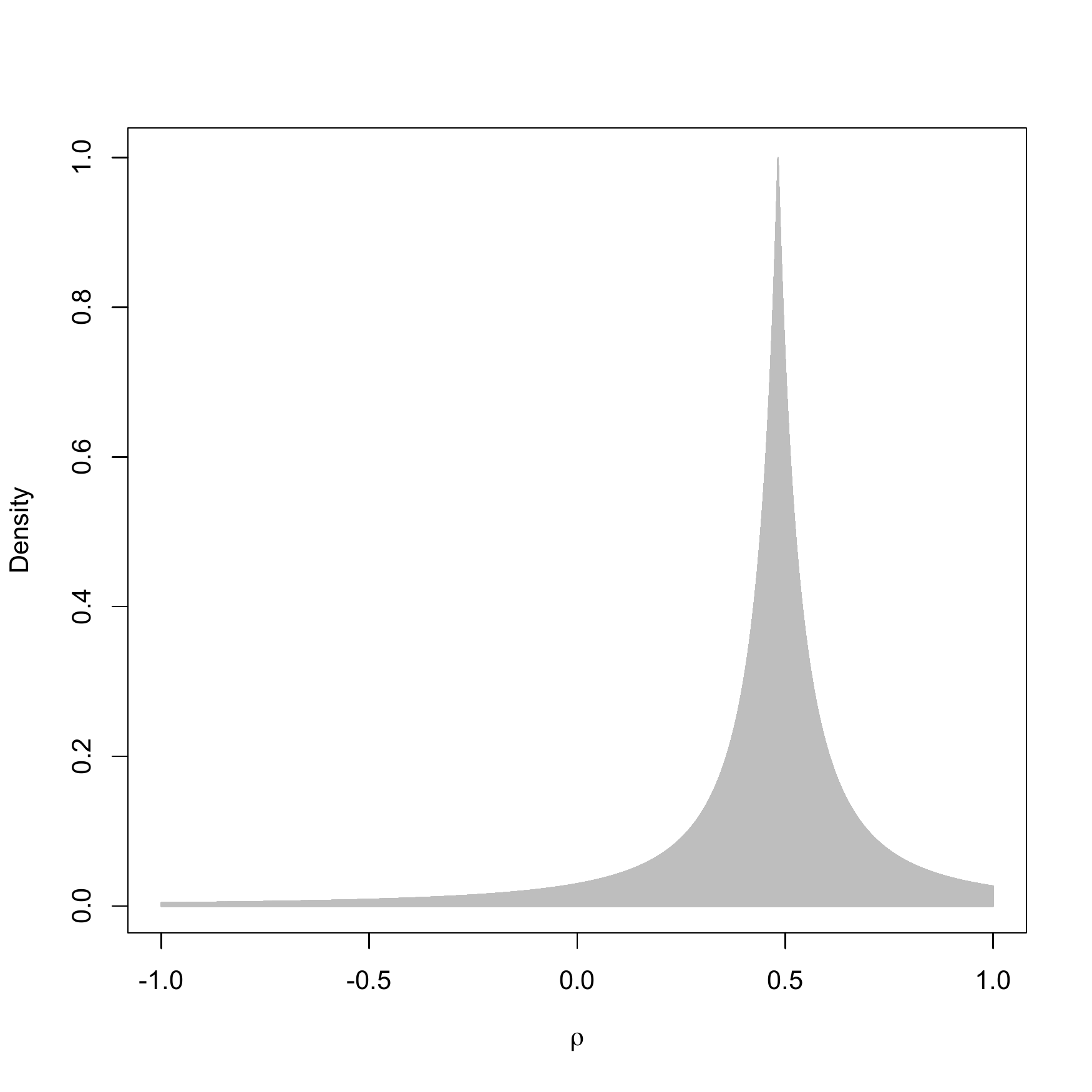}
	\caption{Approximation of the posterior distribution of the Spearman's $\rho$ as defined in \eqref{eq:rho} for the data described in Figure \ref{fig:copula_gen}.}
	\label{fig:rho_post}
\end{figure}

Figure \ref{fig:rho_post} described the approximation to the posterior distribution of $\rho$, as defined in \eqref{eq:multi_rho}, obtained via Algorithm \ref{BCOP}: it is possible to see that the posterior distribution is concentrated around the true value from which the data have been generated. 

The {\tt R} code used is available in the Appendix (``Copula code'').

As noted above, one of the key considerations in developing and implementing $BC_{el}$ is the choice of constraints for the EL. 
This consideration is not particular to $BC_{el}$, but applies to all EL methods. However, the difference here 
is that the selected constraints must be also applicable to the ABC context. With this goal in mind, \citet{Ruli2015} advocate the use 
of scaled composite likelihood score functions as summary statistics in ABC. The scaling takes into account a measure of the relative 
amount of information provided by the different parameters. They argue that the corresponding ABC procedure is therefore invariant to 
reparametrisation and accommodates automatically the curvature of the posterior distribution. This approach is argued to be an 
improvement over that proposed by \citet{Pauli2011} and more ‘fully ABC’ than the $BC_{el}$ approach.

\section*{Acknowledgements}

CCD was supported by an Australian Research Council's Discovery Early Career Researcher Award funding scheme (DE160100741).  KM was supported by the Australian Research Council.  CCD is an Associate Investigator and KM is a Chief Investigator of the Australian Centre of Excellence for Mathematical and Statistical Frontiers (ACEMS).

\bibliographystyle{apalike} 
\bibliography{refsabc010715}

\newpage

\section*{Appendix}

\subsection*{Myeloma code}

\begin{verbatim}
data(myeloma)
survtimes <- myeloma[,1]    # survival times
censtatus <- myeloma[,2]    # vital status (0=alive, 1=dead)
myfun1 <- function(t){ as.numeric(t <= 10) } 
el.cen.EM(fun=myfun1, x=survtimes, d=censtatus, mu=0.2)  
\end{verbatim}

\subsection*{Bayes factor code}

\begin{verbatim}
# test Model 1 (A,B,g,k)=(0,1,1,0) [skew] 
# versus Model 2 (0,1,0.5,0) [less skew] and 
# Model 3 (0,1,0,0) [standard normal]
# Compare B12=el_1/el_2 and B13=el1/el3
# Two sample sizes: n=100, 1000
library(emplik)
# set qc; traditionally set at 0.8 
qc=0.8
# specify the models of interest; qp1 is the 'true' model
qp1=c(0,1,1,0) ;  qp2=c(0,1,0.5,0) ; qp3=c(0,1,0,0)
# specify the quantiles for each model  
refp=c(0.1,0.25,0.5,0.75,0.9)
simq1=qp1[1]+qp1[2]*(1+qc*((1-exp(-qp1[3]*refp))/
(1+exp(qp1[3]*refp))))*((1+refp^2)^qp1[4])*refp
simq2=qp2[1]+qp2[2]*(1+qc*((1-exp(-qp2[3]*refp))/
(1+exp(qp2[3]*refp))))*((1+refp^2)^qp2[4])*refp
simq3=qp3[1]+qp3[2]*(1+qc*((1-exp(-qp3[3]*refp))/
(1+exp(qp3[3]*refp))))*((1+refp^2)^qp3[4])*refp
# set sample size
nob=c(100, 500) 	# no. observations
lennob=length(nob)
nrep=100        # replicates of BF12
# set up matrices and vectors
BF12=logBF12=BF13=logBF13=matrix(0,nrep,lennob)
th1=th2=th3=rep(0,nrep)
# compute BF using el.test based on true parameters for M1 vs M2, M3 
for (nk in 1:lennob){
dth1=dth2=dth3=matrix(0,nrow=nob[nk],ncol=length(refp))
for (repk in 1:nrep){
# generate reference data
zp=qnorm(runif(nob[nk]))
dob=qp1[1]+qp1[2]*(1+qc*((1-exp(-qp1[3]*zp))/
(1+exp(-qp1[3]*zp))))*((1+zp^2)^qp1[4])*zp         
for (k in 1:nob[nk]){
for (j in 1:length(refp)){
dth1[k,j] = (dob[k]<simq1[j])*1 
dth2[k,j] = (dob[k]<simq2[j])*1 
dth3[k,j] = (dob[k]<simq3[j])*1
}}
th1=el.test(dth1,mu=refp)
th2=el.test(dth2,mu=refp)
th3=el.test(dth3,mu=refp)
thll1=th1$'-2LLR' ;  thll2=th2$'-2LLR' ;  thll3=th3$'-2LLR'
logBF12[repk,nk] = -0.5*(thll1 - thll2) 
logBF13[repk,nk] = -0.5*(thll1 - thll3)
BF12[repk,nk] = exp(logBF12[repk,nk])
BF13[repk,nk] = exp(logBF13[repk,nk])
}}  #end of repk, nk
par(mfrow=c(2,2))
logBF123=cbind(logBF12,logBF13)
boxplot(logBF123[,1:4],ylim=c(-10,50),
xlab="1=M1 v M2,n=100; 2=M1 v M2,n=500; 3=M1 v M3, n=100; 4=M1 v M3, n=500", 
ylab="log BF")
\end{verbatim}

\subsection*{Copula code}

\begin{verbatim}
### Function to generate from a quantile function

quantile.fun=function(z,A,B,g,k,c=0.8)
{
val = A + B * ( 1 + c * (1-exp(-g*z)) / (1+exp(-g*z)) ) * 
( 1+z^2 )^k * z
return(val)
}

### Simulations from the copula with a fixed Spearman's rho

# Generation from the copula
library(copula)
cc=claytonCopula(d=5,param=1.076)
uu=rCopula(1000,cc)

# Generation from the normal
z=matrix(NA,nrow=1000,ncol=5)
for(i in 1:5)
{
z[,i]=qnorm(uu[,i])
}

# Generation from the quantile distribution
quant.sim=matrix(NA,nrow=1000,ncol=5)
for(i in 1:5)
{
quant.sim[,i]=quantile.fun(z[,i],A=0,B=1,g=0.5,k=0)
}

#### (Nonparametric) estimation of the marginals

n=1000
F.hat=matrix(NA,nrow=1000,ncol=5)
for(i in 1:5)
{
for(j in 1:1000)
{
F.hat[j,i]=sum(quant.sim[,i]<quant.sim[j,i])/n
}
}

#### BCOP for the Spearman's rho

n=dim(F.hat)[1] 
d=dim(F.hat)[2]
S=10^5

# Ranks
U.hat=matrix(NA,ncol=d,nrow=n)
for(i in 1:d)
{
U.hat[,i]=rank(F.hat[,i])/n
}
VV1=apply(1-U.hat,1,prod)
VV2=apply(U.hat,1,prod)

# Frequentist estimate
const=(d+1)/(2^d-(d+1))
estim1=const*(2^d/n*sum(VV1)-1)

# BCOP 

rho=runif(S, -1,1)
omega=rep(0,S)

for (s in 1:S)
{
est=estim1 - rho[s]
omega1[s]<-exp(-EL(est)$elr)
}

rho.sim=cbind(rho, omega)

plot(rho.sim[,1],rho.sim[,2],type="h",
xlab=expression(rho),ylab="Density",main="",col="grey")

par(mfrow=c(1,3))
plot(uu[,1],uu[,2],xlab=expression(u[1]),ylab=expression(u[2]),
main="Uniform data")
plot(z[,1],z[,2],xlab=expression(z[1]),ylab=expression(z[2]),
main="Normal data")		
plot(quant.sim[,1],quant.sim[,2],xlab=expression(Q(z[1](p))),
ylab=expression(Q(z[2](p))),main="Data from quantile distributions")
\end{verbatim}

\end{document}